\documentclass[12pt,paper,onecolumn]{IEEEtran}

\usepackage{amsmath}
\usepackage{amssymb}
\usepackage{latexsym}
\usepackage{amsfonts}

\begin{document}

\pagestyle{headings}  
\vspace{1cm}

%

\title{  Structure Analysis  on the $k$-error Linear Complexity for $2^n$-periodic Binary Sequences   }

\author{
\authorblockN{Jianqin Zhou\\}
\authorblockA{ Department of Computing, Curtin University, Perth, WA 6102 Australia\\
 School of Computer Science, Anhui Univ. of
Technology, Ma'anshan, 243032 China\\ \ \ zhou9@yahoo.com\\
\ \\
Wanquan Liu\\
Department of Computing, Curtin University, Perth, WA 6102 Australia\\
 w.liu@curtin.edu.au\\
 \ \\
Xifeng Wang\\
School of Computer Science, Anhui Univ. of
Technology, Ma'anshan, 243032 China\\
 \ \\             }
        }
\maketitle              

\begin{abstract}
In this paper, in order to characterize the critical error linear
complexity spectrum (CELCS) for $2^n$-periodic binary sequences, we
first propose a decomposition based on the cube theory. Based on the
proposed $k$-error cube decomposition, and the famous
inclusion-exclusion principle, we obtain the complete
characterization of $i$th descent point (critical point) of the
k-error linear complexity for $i=2,3$. Second, by using the sieve
method and Games-Chan algorithm,  we characterize the
 second descent point (critical point) distribution of the $k$-error linear complexity for $2^n$-periodic binary
sequences. As a consequence, we obtain the complete counting
functions on the $k$-error linear complexity of $2^n$-periodic
binary sequences as the  second descent point  for $k=3,4$. This is
the first time for the second and the third descent points to be
completely characterized. In fact, the proposed constructive
approach has the potential to be used for constructing
$2^n$-periodic binary sequences with the given linear complexity and
$k$-error linear complexity (or CELCS), which is a challenging
problem to be deserved for further investigation in future.


\end{abstract}

\section{Introduction}

The linear complexity of a sequence $s$, denoted as $L(s)$, is
defined as the length of the shortest linear feedback shift register
(LFSR) that can generate $s$. According to the Berlekamp-Massey
algorithm \cite{Massey}, if the linear complexity of a sequence $s$
is $L(s)$, and $2L(s)$ consecutive elements of the sequence are
known, then the whole sequence can be determined. So the linear
complexity of a key sequence should be large enough to resist known
plain text attack. As a measure on the stability of linear
complexity for sequences, the weight complexity and sphere
complexity were defined in the monograph by Ding, Xiao and Shan in
1991 \cite{Ding}. Similarly, Stamp and Martin \cite{Stamp}
introduced the $k$-error linear complexity, which is in essence the
same as the sphere complexity. Specifically, suppose that $s$ is a
sequence with period $N$. For any $k(0\le k\le N)$, the $k$-error
linear complexity of $s$, denoted as $L_k(s)$,  is defined as the
smallest linear complexity that can be obtained when any $k$ or
fewer elements of the sequence are changed within one period.

The reason why people study the stability of linear complexity is
that a small number of element changes may lead to a sharp decline
of linear complexity. How many elements have to be changed to reduce
the linear complexity? Kurosawa et al. in \cite{Kurosawa} introduced
the concept of minimum error($s$) to deal with the problem, and
defined it as the minimum number $k$ for which the $k$-error linear
complexity is strictly less than the linear complexity of sequence
$s$, which is determined by $2^{W_H(2^n-L(s))}$, where $W_H(a)$
denotes the Hamming weight of the binary representation of an
integer $a$.
 In \cite{Meidl}, for the period length $p^n$, where $p$ is an odd prime
and 2 is a primitive root modulo $p^2$, a relationship is
established between the linear complexity and the minimum value $k$
for which the $k$-error linear complexity is strictly less than the
linear complexity. In \cite{Zhou}, for sequences over $GF(q)$ with
period $2p^n$, where $p$ and $q$ are odd primes, and $q$ is a
primitive root modulo $p^2$, the minimum value $k$ is presented for
which the $k$-error linear complexity is strictly less than the
linear complexity.

In another research direction, Rueppel \cite{Rueppel} derived  the number of $2^n$-periodic binary
sequences with given linear complexity $L, 0\le L \le 2^n$. For
$k=1,2$, Meidl \cite{Meidl2005} characterized the complete counting
functions on the $k$-error linear complexity of $2^n$-periodic
binary sequences with linear complexity $2^n$.
For $k=2,3$, Zhu and
Qi \cite{Zhu} further gave the complete counting functions on the
$k$-error linear complexity of $2^n$-periodic binary sequences with
linear complexity $2^n-1$. By using algebraic and combinatorial
methods, Fu et al. \cite{Fu} characterized the
 $2^n$-periodic binary sequences with the $1$-error linear
 complexity and derived the counting function completely for the $1$-error
 linear complexity of  $2^n$-periodic binary sequences.
The complete
counting functions for the number of $2^n$-periodic binary sequences
with the $3$-error linear complexity  are characterized recently in
\cite{Zhou_Liu}.

The CELCS (critical error linear complexity
spectrum) is  studied in \cite{Lauder, Etzion}. The CELCS of a
sequence $s$ consists of the ordered set of points $(k,L_k(s))$
satisfying $L_k(s)> L_{k'}(s)$, for $k'>k$.
In fact they are the points
where a decrease occurs for the $k$-error linear complexity, and thus
are called critical points.

Kurosawa et al. in \cite{Kurosawa} gave an important result about
first descent point of the k-error linear complexity. Due to its
difficulty, the second descent point is rarely investigated in
literature. In this paper, we propose a $k$-error cube decomposition
for $2^n$-periodic binary sequences to investigate the $i$th descent
point of the k-error linear complexity. By applying the famous
inclusion-exclusion principle in combinatorics, we obtain the
complete characterization of $i$th descent point of the k-error
linear complexity for $i=2,3$.

One of our main results is that there exists a unique $k$-error cube
decomposition for a given $2^n$-periodic binary sequence. With a
given series of linear complexity values
$L(c^{(0)}),L(c^{(1)}),L(c^{(2)}), \cdots, L(c^{(m)})$, our focus is
how to construct a sequence $s^{(n)}$ with the right $k$-error cube
decomposition $s^{(n)}=c^{(0)}+c^{(1)}+c^{(2)}+\cdots+c^{(m)}$,  so
that $L(c^{(i)})=L^{(i)}(s^{(n)})$, where $L^{(i)}(s^{(n)})$ is the $k$-error linear complexity of the
$i$th descent point for  $s^{(n)}$.

Next we present a constructive approach for characterizing CELCS for
$2^n$-periodic binary sequences based on the idea reported in
\cite{Zhou_Liu}. Accordingly, the
 second descent point (critical point) distribution of the $k$-error linear complexity for $2^n$-periodic binary
sequences  is characterized. As a consequence, we obtain the complete counting
functions on the $k$-error linear complexity as the  second descent point of
$2^n$-periodic binary sequences   for $k=3,4$.
We expect that with the constructive approach proposed here, one can further obtain other second and third descent point  distribution of the $k$-error linear complexity for $2^n$-periodic binary
sequences.

In \cite{Zhou_Liu}, we investigate all $2^n$-periodic binary
sequences with the given $3$-error linear complexity. In contrast,
here we only investigate the $2^n$-periodic binary sequence with the
given $3$-error linear complexity, where the second decrease occurs
exactly at $3$-error linear complexity. So the result here is more
accurate. It is known by  Kurosawa et al.  \cite{Kurosawa} that for
a $2^n$-periodic binary sequence with  linear complexity $2^n
-(2^i+2^j), 0\le i<j<n$, $4$-error linear complexity is the first
descent point.  However, here we will characterize
 $2^n$-periodic binary sequences with  $4$-error linear complexity as the second descent point, which is a more complex case.

In previous research, investigators focus on the linear complexity
and $k$-error complexity for a given sequence. In this paper, the
motivation of this paper  is to construct $2^n$-periodic binary
sequences with the given linear complexity and $k$-error linear
complexity (or CELCS), and this is a more challenging problem with
broad applications.

The rest of this paper is organized as follows. In  Section II, we
first give  an outline about our main approach  for characterizing  CELCS for $2^n$-periodic binary
sequences. Also some preliminary results are presented.
In Section III, the  $k$-error cube decomposition for $2^n$-periodic binary
sequences  is proposed to investigate the  $i$th descent point of the k-error linear complexity.
By applying the famous   inclusion-exclusion principle, the complete characterization of $i$th descent point of the k-error linear complexity  is presented for $i=2,3$.
 In Section IV, the second descent point (critical point) distribution of the $3$-error
linear complexity for $2^n$-periodic binary sequences is
characterized and also the complete counting functions on the
$3$-error linear complexity  is presented. In Section V, the
 second descent point  distribution of the $4$-error linear complexity for $2^n$-periodic binary
sequences is  characterized and also the complete counting functions on the $4$-error linear complexity as the  second descent point is presented.
   Concluding remarks are  given in Section VI.

\section{Preliminaries}

In this section we first give some preliminary results which will be
used in the sequel. At the same time an outline about the proposed
constructive approach is presented for characterizing  CELCS  for the $k$-error
linear complexity distribution of $2^n$-periodic binary sequences.

 Let $x=(x_1,x_2,\cdots,x_n)$ and
$y=(y_1,y_2,\cdots,y_n)$ be vectors over $GF(q)$. Then define
$x+y=(x_1+y_1,x_2+y_2,\cdots,x_n+y_n)$.
When $n=2m$, we define $Left(x)=(x_1,x_2,\cdots,x_m)$ and
$Right(x)=(x_{m+1},x_{m+2},\cdots,x_{2m})$.

The Hamming weight of an $N$-periodic sequence $s$ is defined as the
number of nonzero elements in per period of $s$, denoted by
$W_H(s)$. Let $s^N$ be one period of $s$. If $N=2^n$, $s^N$ is also
denoted as $s^{(n)}$.
The absolute distance of two elements is defined as the difference of their indexes.

The linear complexity of a $2^n$-periodic binary sequence $s$
  can be recursively computed by the Games-Chan algorithm \cite{Games} stated as follows.

\noindent {\bf Algorithm  2.1}

\noindent {\bf Input}: A $2^n$-periodic binary sequence $s=[Left(s),Right(s)]$, $c=0$.

\noindent {\bf Output}:  $L(s)=c$.

\noindent Step 1. If $Left(s)=Right(s)$, then deal with $Left(s)$ recursively. Namely, $L(s)=L(Left(s))$.

\noindent Step 2. If $Left(s)\neq Right(s)$, then $c=c+2^{n-1}$ and deal with $Left(s)\bigoplus Right(s)$ recursively.
Namely, $L(s)=2^{n-1}+L(Left(s)\bigoplus Right(s))$.

Repeat Step 1 and Step 2 recursively until one element is left.

\noindent Step 3. $s=(a)$,  if $a=1$ then $c=c+1$, else $c=c+0$.

\

The following two lemmas are  well known results on $2^n$-periodic
binary sequences. Please refer to \cite{Meidl2005,Zhou_Liu,Zhu} for details.

\noindent {\bf Lemma  2.1} Suppose that $s$ is a binary sequence with
period $N=2^n$. Then $L(s)=N$ if and only if the Hamming weight of a
period of the sequence is odd.

\noindent {\bf Lemma 2.2}  Let $s_1$ and $s_2$ be two binary sequences
with period $2^n$. If $L(s_1)\ne L(s_2)$, then
$L(s_1+s_2)=\max\{L(s_1),L(s_2)\} $; otherwise if $L(s_1)= L(s_2)$,
then $L(s_1+s_2)<L(s_1)$.

Suppose that the linear complexity of $s$ can decrease when at least
$k$ elements of $s$ are changed. By Lemma 2.2, the linear complexity
of the binary sequence, in which elements at exactly those $k$
positions are all nonzero, must be $L(s)$. Therefore, for the
computation of the $k$-error linear complexity, we only need to find
the binary sequence whose Hamming weight is the minimum and its
linear complexity is $L(s)$.

Based on Games-Chan algorithm, the following  lemma is given in
\cite{Meidl2005}.

\noindent {\bf Lemma  2.3} Suppose that $s$ is a binary sequence
with one period $s^{(n)}=\{s_0,s_1,s_2,\cdots, s_{2^n-1}\}$, a
mapping $\varphi_n$ from $F^{2^n}_2$ to $F^{2^{n-1}}_2$ is defined
as
\begin{eqnarray*}\varphi_n(s^{(n)})
&=&\varphi_n((s_0,s_1,s_2,\cdots,
s_{2^n-1}))\\
&=&(s_0+s_{2^{n-1}},s_1+s_{2^{n-1}+1},\cdots,
s_{2^{n-1}-1}+s_{2^n-1}) \end{eqnarray*}

Let $W_H(\mathbf{\upsilon})$ denote the Hamming weight of a vector
$\mathbf{\upsilon}$. Then the mapping $\varphi_n$ has the following
properties.

1) $W_H(\varphi_n(s^{(n)}))\le W_H(s^{(n)})$;

2) If $n\ge2$, then $W_H(\varphi_n(s^{(n)}))$ and $W_H(s^{(n)})$ are
either both odd or both even;

3) The set $$\varphi^{-1}_{n+1}(s^{(n)})=\{v\in
F^{2^{n+1}}_2|\varphi_{n+1}(v)=s^{(n)} \}$$ of the preimage of
$s^{(n)}$ has cardinality $2^{2^n}$.

Rueppel \cite{Rueppel} presented the following preliminary result on
the number of sequences with a given linear complexity.

\noindent {\bf Lemma  2.4}  The number $N(L)$ of $2^n$-periodic
binary sequences with linear complexity $L, 0\le L \le 2^n$, is
given by $N(L)=\left\{\begin{array}{l}
1, \ \ \ \ \ L=0\ \   \\
2^{L-1}, \ 1\le L\le 2^n
\end{array}\right.$\

\

Based on algebraic and  combinatorial methods, Fu et al. \cite{Fu} characterized the
 $2^n$-periodic binary sequences with the $1$-error linear
 complexity and derived the counting function completely for the $1$-error
 linear complexity of  $2^n$-periodic binary sequences.
 Meidl \cite{Meidl2005} characterized the complete counting
functions on the $1$-error linear complexity of $2^n$-periodic
binary sequences with linear complexity $2^n$.  Zhu and
Qi \cite{Zhu}  gave the complete counting functions on the
$2$-error linear complexity of $2^n$-periodic binary sequences with
linear complexity $2^n-1$.

In this paper, in order to characterize CELCS (critical error linear
complexity spectrum), we will use the {\it Cube Theory} recently
introduced in \cite{Zhou_Liu2013}.  Cube theory and some related
results are presented next for completeness.



Suppose that the position difference
 of two non-zero elements of a sequence $s$ is $(2x+1)2^y$,
where  $x$ and $y$ are non-negative integers. From Algorithm 2.1,
only in the $(n-y)$th step, the  sequence length  is $2^{y+1}$, so
the two non-zero elements   must be  in the left and right half of
the sequence respectively, thus they can be removed or reduce to one
non-zero element in consequence operation. Therefore we have the
following definitions.

 {\bf Definition  2.1} (\cite{Zhou_Liu2013}) Suppose that the position difference of
  two non-zero elements of a sequence $s$ is $(2x+1)2^y$,
where both $x$ and $y$ are non-negative integers, then the distance
between the two elements is defined as  $2^y$.

 {\bf Definition  2.2} (\cite{Zhou_Liu2013}) Suppose that $s$ is a binary sequence
with period $2^n$, and there are $2^m$ non-zero elements in $s$, and
$0\le i_1< i_2<\cdots<i_m<n$. If $m=1$, then there are 2 non-zero
elements in $s$ and the distance between the two elements is
$2^{i_1}$, so it is called as a 1-cube. If $m = 2$, then $s$ has 4
non-zero elements which form a rectangle, the lengths of 4 sides are
$2^{i_1}$ and $2^{i_2}$ respectively, so it is called as a 2-cube.
In general, $s$ has $2^{m-1}$ pairs of non-zero elements, in which
there are $2^{m-1}$ non-zero elements which form a $(m-1)$-cube, the
other $2^{m-1}$ non-zero elements also form a $(m-1)$-cube, and the
distance between each pair of elements are all  $2^{i_m}$, then the
sequence $s$ is called as an $m$-cube, and the linear complexity of $s$ is
 called as the linear complexity of the cube as well.

 {\bf Definition  2.3} (\cite{Zhou_Liu2013})
 A non-zero element  of sequence $s$ is  called a vertex.
Two vertexes can  form an edge. If the distance between the two
elements (vertices) is   $2^y$, then the length of the edge is
defined as $2^y$.

As demonstrated in \cite{Zhou_Liu2013}, the linear complexity of a
$2^n$-periodic binary sequence with only one cube has the following
nice property.

 {\bf Theorem   2.1} Suppose that $s$ is a binary sequence
with period $2^n$, and non-zero elements of $s$ form a $m$-cube, if
lengths of  edges are $ i_1, i_2,\cdots ,i_m$ $(0\le i_1<
i_2<\cdots<i_m<n )$ respectively, then
$L(s)=2^n-(2^{i_1}+2^{i_2}+\cdots+2^{i_m})$.

\begin{proof}\
 we give a proof based on Algorithm 2.1.

In the $k$th step, $1\le k\le n$, if and only if one period of the sequence can not be divided into two equal parts, then the
 linear complexity should be increased by half period. In the $k$th step, the
 linear complexity can be increased by maximum $2^{n-k}$.

Suppose that non-zero elements of sequence $s$ form a $m$-cube,
lengths of  edges are $ i_1, i_2,\cdots ,i_m$ $(0\le i_1<
i_2<\cdots<i_m<n )$ respectively. Then in the $(n-i_m)$th step, one
period of the sequence can  be divided into two equal parts, then
the
 linear complexity should not be increased by $2^{i_m}$.

$\cdots\cdots$

 In the $(n-i_2)$th step, one period of the sequence can  be divided into two equal parts, then the
 linear complexity should not be increased by $2^{i_2}$.

 In the $(n-i_1)$th step, one period of the sequence can  be divided into two equal parts, then the
 linear complexity should not be increased by $2^{i_1}$.

Therefore, $L(s)=1+1+2+2^2+\cdots+2^{n-1}-(2^{i_1}+2^{i_2}+\cdots+2^{i_m})=2^n-(2^{i_1}+2^{i_2}+\cdots+2^{i_m})$.

The proof is complete now.
\end{proof}\

Based on  Algorithm 2.1, we may have a standard cube decomposition
for any binary sequence with period $2^n$.

\noindent {\bf Algorithm 2.2}

\noindent {\bf Input:} $s^{(n)}$ is a binary sequence with period
$2^n$.

\noindent {\bf Output:} A  cube decomposition of sequence $s^{(n)}$.

\noindent Step 1. Let $s^{(n)}=[Left(s^{(n)}),Right(s^{(n)})]$.

\noindent Step 2. If $Left(s^{(n)})=Right(s^{(n)})$, then we only consider $Left(s^{(n)})$.

\noindent Step 3. If $Left(s^{(n)})\neq Right(s^{(n)})$, then we
consider $Left(s^{(n)})\bigoplus Right(s^{(n)})$.  In this case, some  nonzero elements of  $s$
may be removed.

\noindent Step 4. After above operation, we can obtain one nonzero element. Now by only restoring the  nonzero elements in $Right(s^{(n)})$ removed in Step 2,
so that $Left(s^{(n)})=Right(s^{(n)})$. In this case, we obtain a cube $c_1$ with linear complexity $L(s^{(n)})$.

\noindent Step 5. With $s^{(n)}\bigoplus c_1$, run Step 1 to Step 4. We obtain a cube $c_2$ with linear complexity less than $L(s^{(n)})$.

\noindent Step 6. With these nonzero elements  left in $s^{(n)}$, run Step 1 to Step 5 recursively
 we will obtain
a series of cubes in the descending order of linear complexity.

\

 Obviously, this is a cube decomposition of sequence $s^{(n)}$, and we define it as {\bf
the standard cube decomposition}. One can observe that cube
decomposition of a sequence may not be unique in general and {\bf
the standard cube decomposition} of a sequence described above is
unique.

Next we use  a sequence $\{1101\ 1001\ 1000\ 0000\}$ to illustrate the decomposition process.


As $Left\neq Right$, then we consider $Left\bigoplus Right$. Then
the cube $\{1000\ 0000\ 1000\ 0000\}$ is removed.

Recursively, as $Left\neq Right$, then we consider $Left\bigoplus
Right$. This time the cube $\{0001\ 0001\ 0000\ 0000\}$ is removed. Only cube $\{0100\ 1000\ 0000\ 0000\}$
is retained. So the standard cube decomposition
 is $\{0100\ 1000\ 0000\ 0000\}$,  $\{0001\ 0001\ 0000\ 0000\}$,  $\{1000\ 0000\ 1000\ 0000\}$.

\section{A constructive approach for computing   descent points of the
k-error linear complexity}

How many elements have to be changed to decrease the linear complexity? For a $2^n$-periodic binary
sequence $s^{(n)}$, Kurosawa et al. in \cite{Kurosawa}
 showed that the first descent point  of the $k$-error linear complexity is
 reached by
$k=2^{W_H(2^n-L(s^{(n)}))}$, where $W_H(a)$ denotes the Hamming
weight of the binary representation of an integer $a$.

In this section, first,  the  $k$-error cube decomposition of
$2^n$-periodic binary sequences is developed based on the proposed
cube theory.  Second we
 investigate the formula to determine the second  descent point for the
k-error linear complexity of  $2^n$-periodic binary
sequences based on the linear complexity and the first descent points for the
k-error linear complexity. Third
 we  study the formula to determine the third  descent points for the
k-error linear complexity  based on the linear complexity, the first and second descent points for the
k-error linear complexity.

For clarity of presentation, we first introduce some definitions.

Let $k^{(i)}$ denote the $i$th descent point  of the $k$-error
linear complexity, where $i>0$. We define $S(a)$ as the binary
representation of an integer $a$, and $W_H(S(a))$ denotes the
Hamming weight of $S(a)$. We further define $L^{(i)}(s^{(n)})$ as
the $k$-error linear complexity of the $i$th descent point for a
$2^n$-periodic binary sequence $s^{(n)}$, and define
$$S(s^{(n)})=S(2^n-L(s^{(n)})) $$
$$S^{(i)}(s^{(n)})=S(2^n-L^{(i)}(s^{(n)})) $$
where $i\ge0$ and $L(s^{(n)})$ is also denoted as
$L^{(0)}(s^{(n)})$. For a given binary digit representation $S_1$,
one can prove easily that there exists only one linear complexity
value $L_1=2^n-(2^{i_1}+2^{i_2}+\cdots+2^{i_m})$, where $0\le i_1<
i_2<\cdots<i_m<n$, such that $S_1=S(2^n-L_1)$. In this case, we
define
$$S^{-1}(S_1)=i_1, S^{-m}(S_1)=i_m$$
$$S_{>i_k}(2^n-L_1)=S(2^{i_{k+1}}+2^{i_{k+2}}+\cdots+2^{i_m})$$

 Let $S(a)=(x_1,x_2,\cdots,x_n)$ and
$S(b)=(y_1,y_2,\cdots,y_n)$. Then define $S(a)\cap S(b)=(x_1 y_1,x_2
y_2,\cdots,x_n  y_n)$,  $S(a)\cup S(b)=(x_1+y_1-x_1 y_1,x_2+y_2-x_2
  y_2,\cdots,x_n+y_n-x_n y_n)$.

Next we present a very fundamental theorem regarding CELCS, followed by  an  important  definition called {\bf the $k$-error cube decomposition}.

\noindent {\bf Theorem 3.1} Let $s^{(n)}$ be a $2^n$-periodic binary
sequence. Then

i) $s^{(n)}=c^{(0)}+c^{(1)}+c^{(2)}+\cdots+c^{(j)}$, where $c^{(i)}$ is a cube with  linear complexity $L(c^{(i)})=L^{(i)}(s^{(n)})$ and $k^{(i+1)}= W_H(c^{(0)}+c^{(1)}+c^{(2)}+\cdots+c^{(i)}), 0\le i\le j$;

ii) $s^{(n)}=c^{(0)}+c^{(1)}+c^{(2)}+\cdots+c^{(m)}+t_m^{(n)}$, where
 $c^{(i)}$ is a cube with  linear complexity $L(c^{(i)})=L^{(i)}(s^{(n)})$,
  $t_m^{(n)}$ is a $2^n$-periodic binary
sequence with $L^{(m)}(s^{(n)})>L(t_m^{(n)})$, and $k^{(m+1)}\le W_H(c^{(0)}+c^{(1)}+c^{(2)}+\cdots+c^{(m)})$.

\begin{proof}\ i) Suppose that the last decent point of $k$-error linear complexity is $(k^{(j+1)},L^{(j+1)}(s^{(n)}))$. Then $L^{(j+1)}(s^{(n)})=0$.
Assume that the second last decent point of $k$-error linear complexity is $(k^{(j)},L^{(j)}(s^{(n)}))$, and $L^{(j)}(s^{(n)})$ is achieved with a cube $c^{(j)}$ by adding a $2^n$-periodic binary
sequence $e_j^{(n)}$ to $s^{(n)}$, where $k^{(j)}=W_H(e_j^{(n)})$. Thus $e_j^{(n)}+s^{(n)}=c^{(j)}$, which implies that $s^{(n)}=e_j^{(n)}+c^{(j)}$.

By the definition of $k$-error linear complexity, $W_H(e_j^{(n)})<W_H(s^{(n)})$.

Similarly, $e_j^{(n)}=e_{j-1}^{(n)}+c^{(j-1)}$, and $W_H(e_{j-1}^{(n)})<W_H(e_j^{(n)})$. 
If $L(c^{(j-1)})\le L(c^{(j)})$, as $s^{(n)}=e_j^{(n)}+c^{(j)}=e_{j-1}^{(n)}+c^{(j-1)}+c^{(j)}$, then
adding a $2^n$-periodic binary
sequence $e_{j-1}^{(n)}$ to $s^{(n)}$, in this case $L(e_{j-1}^{(n)}+s^{(n)})<L(c^{(j)})$.
This contradicts the fact that $k^{(j)}=W_H(e_j^{(n)})$. Thus $L(c^{(j-1)})> L(c^{(j)})$ and $k^{(j-1)}=W_H(e_{j-1}^{(n)})$.

$\cdots \cdots$

Finally, based on the above analysis, we have that $s^{(n)}=c^{(0)}+c^{(1)}+c^{(2)}+\cdots+c^{(j)}$, where $L(c^{(0)})>L(c^{(1)})>L(c^{(2)})>\cdots>L(c^{(j)})$, $L^{(i)}(s^{(n)})=L(c^{(i)})$, and $k^{(i)}=W_H(e_{i}^{(n)})= W_H(c^{(0)}+c^{(1)}+c^{(2)}+\cdots+c^{(i-1)}), 0\le i\le j+1$.

\

ii) In the case of i), we first obtain the last cube $c^{(j)}$, then $c^{(j-1)}$, $c^{(j-2)}$, $\cdots \cdots$.

In this case, we first obtain the cube $c^{(m)}$, so that $L^{(m)}(s^{(n)})=L(c^{(m)})$.

Assume that $L^{(m)}(s^{(n)})$ is achieved with a cube $c^{(m)}$ by adding a $2^n$-periodic binary
sequence $e_m^{(n)}$ to $s^{(n)}$, which implies that $s^{(n)}=e_m^{(n)}+c^{(m)}+t_m^{(n)}$, where $L(t_m^{(n)})<L(c^{(m)})$.

By applying the result of i) to $e_m^{(n)}$, we have that
 $s^{(n)}=c^{(0)}+c^{(1)}+c^{(2)}+\cdots+c^{(m)}+t_m^{(n)}$, where
 $L(c^{(i)})=L^{(i)}(s^{(n)})$, $k^{(i)}=W_H(e_{i}^{(n)})= W_H(c^{(0)}+c^{(1)}+c^{(2)}+\cdots+c^{(i-1)}), 0\le i\le m$.

 It is obvious that $k^{(m+1)}\le W_H(c^{(0)}+c^{(1)}+c^{(2)}+\cdots+c^{(m)})$.
\end{proof}

In fact, there indeed exists the case that $k^{(m+1)}<
W_H(c^{(0)}+c^{(1)}+c^{(2)}+\cdots+c^{(m)})$. For example, let \\
$c^{(0)}=\{11001100\ 00000000\ 00000000\ 00000000\}$,
\\
$c^{(1)}=\{10101010\  10101010\  00000000\ 00000000 \}$,
\\
$c^{(2)}=\{11001100\ 11001100\ 11001100\ 11001100\}$,\\
 and  $s^{(5)}=c^{(0)}+c^{(1)}+c^{(2)}$.
Then  $L^{(i)}(s^{(5)})=L(c^{(i)}), 0\le i\le 2$. It is easy to verify that $k^{(3)}=12<W_H(s^{(5)})=16$. This is the case of ii).

Let\\
$c^{(2)}=\{01100110\ 01100110\ 01100110\ 01100110\}$,\\
$c^{(3)}=\{10101010\ 10101010\ 10101010\ 10101010\}$.\\
Then $s^{(5)}=c^{(0)}+c^{(1)}+c^{(2)}+c^{(3)}$,  $L^{(i)}(s^{(5)})=L(c^{(i)}), 0\le i\le 3$.
$k^{(3)}=W_H(c^{(0)}+c^{(1)}+c^{(2)})=12$, $k^{(4)}=W_H(c^{(0)}+c^{(1)}+c^{(2)}+c^{(3)})=16$.
 This is the case of i).

We now still use  the sequence $s^{(4)}=\{1101\ 1001\ 1000\ 0000\}$
to illustrate Theorem 3.1.

Let\\ $c^{(0)}= \{0100\ 1000\ 0000\ 0000\}$,\\
$ t_0^{(4)}=\{1001\ 0001\ 1000\ 0000\}$. Then $s^{(4)}=c^{(0)}+t_0^{(4)}$.

Let\\
 $c^{(0)}= \{0000\ 1100\ 0000\ 0000\}$,\\
  $c^{(1)}=\{0101\ 0101\ 0000\ 0000\}$,\\
   $t_1^{(4)}=\{1000\ 0000\ 1000\ 0000\}$. Then $s^{(4)}=c^{(0)}+c^{(1)}+t_1^{(4)}$.

Let\\
 $c^{(0)}= \{0000\ 0100\ 0000\ 1000\}$,\\
  $c^{(1)}=\{0100\ 0100\ 0001\ 0001\}$,\\
  $c^{(2)}=\{1001\ 1001\ 1001\ 1001\}$,\\
    $t_2^{(4)}=\{0000\ 0000\ 0000\ 0000\}$. Then $s^{(4)}=c^{(0)}+c^{(1)}+c^{(2)}+t_2^{(4)}$. It is easy to verify that $k^{(3)}=W_H(s^{(4)})=6$.

\

One can see for any $2^n$-periodic binary sequence $s^{(n)}$, there
is $m>0$, such that  $L^{(m+1)}(s^{(n)})=0$. Then from part one of
Theorem 3.1, we have that
$s^{(n)}=c^{(0)}+c^{(1)}+c^{(2)}+\cdots+c^{(m)}$, where $c^{(0)}$ is
a cube with  linear complexity $L^{}(s^{(n)})$, $c^{(i)}$ is a cube
with $2^{W_H(S^{(i)}(s^{(n)}))}$ nonzero elements and linear
complexity $L^{(i)}(s^{(n)})$, $0<i\le m$.

We define $s^{(n)}=c^{(0)}+c^{(1)}+c^{(2)}+\cdots+c^{(m)}$ as  {\bf
the $k$-error cube decomposition} of  a $2^n$-periodic binary
sequence $s^{(n)}$.
%
For a $2^n$-periodic binary sequence $s^{(n)}$, its $k$-error cube
decomposition may be different from its standard cube decomposition.
For sequence $\{1101\ 1001\ 1000\ 0000\}$ used in standard
decomposition, its $k$-error cube decomposition is different from
its standard decomposition and is given as follows:
 $\{0000\ 0100\ 0000\ 1000\}$, $\{0100\ 0100\ 0001\ 0001\}$, $\{1001\ 1001\ 1001\ 1001\}$.

From part one of Theorem 3.1, one can see that there exists a
$k$-error cube decomposition for a given $2^n$-periodic binary
sequence. Next we will use part two of Theorem 3.1 to find the
second and third descent points.

\noindent {\bf Theorem 3.2} For a $2^n$-periodic binary
sequence $s^{(n)}$, the second descent point  of the $k$-error linear complexity is
 reached by
$k^{(2)}=2^{W_H(S^{}(s^{(n)}))}+2^{W_H(S^{(1)}(s^{(n)}))}-2\times2^{W_H(S(s^{(n)})\cap S^{(1)}(s^{(n)}) )}$.

\begin{proof}\
i) First consider the case that $L(s^{(n)})=2^n$. In this case, $S(s^{(n)})=S(2^n-L(s^{(n)})) $ only contains zero elements.
So, $k^{(2)}=1+2^{W_H(S^{(1)}(s^{(n)}))}-2\times2^0=2^{W_H(S^{(1)}(s^{(n)}))}-1$.

From Theorem 3.1,  $s^{(n)}=c_0+c^{(1)}+t^{(n)}$, where
$c_0$ is a 0-cube
(one nonzero element),  $c^{(1)}$ is a cube with $2^{W_H(S^{(1)}(s^{(n)}))}$ nonzero elements and linear complexity $L^{(1)}(s^{(n)})$, and
$L(t^{(n)})<L^{(1)}(s^{(n)})$.

From Lemma 2.2, $L^{(1)}(s^{(n)})$ is achieved by changing  $c_0$ to a zero element, and $L^{(2)}(s^{(n)})$ is achieved by constructing another cube $c_2$ with linear complexity $L^{(1)}(s^{(n)})$, and using $c_0$ as a nonzero element of  $c_2$.  Thus $k^{(2)}=2^{W_H(S^{(1)}(s^{(n)}))}-1$.

(For example, $u^{(4)}=\{1111\ 1000\ 0000\ 0000 \}$.
$L^{(1)}(u^{(4)})=2^4-(1+2)$ is achieved by a 2-cube  $\{1111\ 0000\
0000\ 0000 \}$. So $L^{(2)}(u^{(4)})$ is achieved by a 3-cube
$\{1111\ 1111\ 0000\ 0000 \}$, $k^{(2)}=2^2-1=3$.)

ii) Second consider the case that $L(s^{(n)})<2^n$.

From Theorem 3.1, suppose that  $s^{(n)}=c^{(0)}+c^{(1)}+t^{(n)}$, where $c^{(0)}$ is a cube with $2^{W_H(S^{}(s^{(n)}))}$ nonzero elements and linear complexity $L^{}(s^{(n)})$,  and   $c^{(1)}$ is a cube with $2^{W_H(S^{(1)}(s^{(n)}))}$ nonzero elements and linear complexity $L^{(1)}(s^{(n)})$, and
$L(t^{(n)})<L^{(1)}(s^{(n)})<L(s^{(n)})$.

If $W_H(c^{(0)}+c^{(1)})=W_H(c^{(0)})+W_H(c^{(1)})$, it is obvious that by changing $2^{W_H(S^{}(s^{(n)}))}-2^{W_H(S(s^{(n)})\cap S^{(1)}(s^{(n)}) )}+2^{W_H(S^{(1)}(s^{(n)}))}-2^{W_H(S(s^{(n)})\cap S^{(1)}(s^{(n)}) )}$ nonzero elements, $L(s^{(n)})$ will be less than $L^{(1)}(s^{(n)})$.

In the case that $W_H(c^{(0)}+c^{(1)})<W_H(c^{(0)})+W_H(c^{(1)})$ and $L(c^{(0)}+c^{(1)})=L(c^{(0)})$, it is easy to show that
$W_H(c^{(0)}+c^{(1)})=2^{W_H(S^{}(s^{(n)}))}-2^{W_H(S(s^{(n)})\cap S^{(1)}(s^{(n)}) )}+2^{W_H(S^{(1)}(s^{(n)}))}-2^{W_H(S(s^{(n)})\cap S^{(1)}(s^{(n)}) )}$.
Thus by changing $2^{W_H(S^{}(s^{(n)}))} +2^{W_H(S^{(1)}(s^{(n)}))}-2\times2^{W_H(S(s^{(n)})\cap S^{(1)}(s^{(n)}) )}$ nonzero elements,  $L(s^{(n)})$ will be less than $L^{(1)}(s^{(n)})$. So $k^{(2)}=2^{W_H(S^{}(s^{(n)}))}+2^{W_H(S^{(1)}(s^{(n)}))}-2\times2^{W_H(S(s^{(n)})\cap S^{(1)}(s^{(n)}) )}$.

(For example, let $c^{(0)}=\{0101\ 0000\ 0000\ 1010 \}$,
$c^{(1)}=\{1010\ 1010\ 1010\ 1010 \}$. Then $c^{(0)}+c^{(1)}=\{1111\
1010\ 1010\ 0000 \}$, where $c^{(0)}$ and $c^{(1)}$ share 2 nonzero
elements $\{1010 \}$. So  $k^{(2)}=2^2+2^3-2\times2^1=8$.)

This completes the proof.

\end{proof}\

In fact, Chang  and Wang proved this result in Theorem 3 of
\cite{Chang},  with a much complicated approach.

\

 Next
 we  investigate the computation of the third  descent point for the
k-error linear complexity  based on the linear complexity, the first
and second descent points for the k-error linear complexity. Before
present our main result, we first give a special result.

\noindent {\bf Proposition 3.1}  For a $2^n$-periodic binary
sequence $s^{(n)}$, let $k^{(i)}$ denote the $i$th descent point  of
the $k$-error linear complexity, $i>0$.  If $
S^{(i)}(s^{(n)})\supset S^{(0)}(s^{(n)})\cup
S^{(1)}(s^{(n)})\cup\cdots\cup S^{(i-1)}(s^{(n)})$,   then
$k^{(i+1)}=2^{W_H( S^{(i)}(s^{(n)}))}-k^{(i)}$, $i>1$.

\begin{proof}\
%
%

 As $ S^{(i)}(s^{(n)})\supset S^{(0)}(s^{(n)})\cup
S^{(1)}(s^{(n)})\cup\cdots\cup S^{(i-1)}(s^{(n)})$, by changing
$2^{W_H( S^{(i)}(s^{(n)}))}-k^{(i)}$ elements of $s^{(n)}$,  the
linear complexity of $s^{(n)}$ becomes 0 or less than
$L^{(i)}(s^{(n)})$.
 So  $k^{(i+1)}=2^{W_H( S^{(i)}(s^{(n)}))}-k^{(i)}$.

\end{proof}

For example, let $s^{(4)}=\{1111\ 1111\ 1110\ 0000\}, n=4$. Then
$S^{(0)}(s^{(n)})=\{0000\}$, $S^{(1)}(s^{(n)})=\{0011\}$,
$S^{(2)}(s^{(n)})=\{0111\}$, $S^{(3)}(s^{(n)})=\{1111\}$. So $
S^{(3)}(s^{(n)})\supset S^{(0)}(s^{(n)})\cup S^{(1)}(s^{(n)})\cup
S^{(2)}(s^{(n)})$.

As
$L^{(1)}(s^{(4)})$ is achieved by a 2-cube  $\{0000\ 0000\ 1111\  0000 \}$, $k^{(1)}=1$,
 $L^{(2)}(s^{(4)})$ is achieved by a 3-cube  $\{1111\ 1111\ 0000\ 0000 \}$, $k^{(2)}=3$. So $k^{(3)}=2^3-3=5$. By changing $k^{(3)}$ elements, $s^{(4)}$ becomes  a 4-cube  $\{1111\ 1111\ 1111\ 1111 \}$.

As $L^{(3)}(s^{(4)})$ is achieved by a 4-cube  $\{1111\ 1111\ 1111\
1111 \}$, $k^{(3)}=5$, thus $k^{(4)}=2^4-5=11$.  By changing
$k^{(4)}$ elements, the linear complexity of $s^{(4)}$ becomes 0.

The above result is for the $i$th descent point computation in some
special cases. Next we will investigate the third descent point in
general. First, we give the the famous principle of
inclusion-exclusion in combinatorics for finite sets $A_1, \cdots,
A_n$, which can be stated as follows.
$$|\bigcup\limits^{n}_{i=1}A_i|=\sum\limits^{n}_{i=1}|A_i|-\sum\limits_{1\le i< j\le n}|A_i\cap A_j|+\sum\limits_{1\le i< j<k\le n}|A_i\cap A_j\cap A_k|-\cdots+(-1)^{n-1}|A_1\cap\cdots\cap A_n|$$

Based on the  principle of inclusion-exclusion, we  give the
following important theorem on the third descent point.

\noindent {\bf Theorem 3.3} For a $2^n$-periodic binary sequence
$s^{(n)}$, let $k^{(i)}$ denote the $i$th descent point  of the
$k$-error linear complexity, $i>0$, and
$i_{m(S1\setminus S0S2)}=S^{-m}\{S^{(1)}(s^{(n)})\setminus [S^{(1)}(s^{(n)})\cap
(S^{(0)}(s^{(n)})\cup S^{(2)}(s^{(n)}))]\}$.

With the following conditions

(i)
   $W_H[S^{(1)}(s^{(n)})\cap
(S^{(0)}(s^{(n)})\cup S^{(2)}(s^{(n)}))]<W_H( S^{(1)}(s^{(n)}))$

(ii) $\{S^{(0)}(s^{(n)})\cap S^{(2)}(s^{(n)})=S^{(0)}(s^{(n)})\cap
S^{(1)}(s^{(n)})\cap S^{(2)}(s^{(n)})$

(iii) $$[i_{m(S1\setminus S0S2)}
>\min\{S^{-1}(S^{(1)}(s^{(n)})\cap
S^{(2)}(s^{(n)})),S^{-1}(S^{(0)}(s^{(n)})\cap S^{(2)}(s^{(n)}))\}$$
and
$$(S_{>i_{m(S1\setminus S0S2)}}^{(0)}(s^{(n)})\cap
S_{>i_{m(S1\setminus S0S2)}}^{(2)}(s^{(n)}))\subset
S^{(1)}(s^{(n)})]\}$$

If (i) and (ii) or (i) and (iii) hold, then \\
$k^{(3)}=2^{W_H(S^{(0)}(s^{(n)}))}+2^{W_H(S^{(1)}(s^{(n)}))}+2^{W_H(S^{(2)}(s^{(n)}))}$
$-2\times2^{W_H(S^{(0)}(s^{(n)})\cap S^{(1)}(s^{(n)})
)}-2\times2^{W_H(S^{(0)}(s^{(n)})\cap S^{(2)}(s^{(n)})
)}-2\times2^{W_H(S^{(1)}(s^{(n)})\cap S^{(2)}(s^{(n)})
)}+2\times2^{W_H(S^{(0)}(s^{(n)})\cap S^{(1)}(s^{(n)})\cap
S^{(2)}(s^{(n)}))}$;

Otherwise, we have

$k^{(3)}=2^{W_H(S^{(0)}(s^{(n)}))}+2^{W_H(S^{(1)}(s^{(n)}))}+2^{W_H(S^{(2)}(s^{(n)}))}-2\times2^{W_H(S^{(0)}(s^{(n)})\cap
S^{(1)}(s^{(n)}) )}-2\times2^{W_H(S^{(0)}(s^{(n)})\cap
S^{(2)}(s^{(n)}) )}-2\times2^{W_H(S^{(1)}(s^{(n)})\cap
S^{(2)}(s^{(n)}) )}+4\times2^{W_H(S^{(0)}(s^{(n)})\cap
S^{(1)}(s^{(n)})\cap S^{(2)}(s^{(n)}))}$.

\


\begin{proof}\
The following  proof is based on the framework that  $s^{(n)}=c^{(0)}+c^{(1)}+c^{(2)}+\cdots+c^{(i)}+t_i^{(n)}$.
 For $c^{(0)}+c^{(1)}+c^{(2)}+\cdots+c^{(i)}$,  by changing $k^{(i+1)}$ elements,
 the linear complexity of $c^{(0)}+c^{(1)}+c^{(2)}+\cdots+c^{(i)}$ can become 0 (in which case  $k^{(i+1)}=W_H(c^{(0)}+c^{(1)}+c^{(2)}+\cdots+c^{(i)}) $) or less than
 $L(c^{(i)})$ (where $k^{(i+1)}<W_H(c^{(0)}+c^{(1)}+c^{(2)}+\cdots+c^{(i)}) $).

 In the case that the linear complexity of $c^{(0)}+c^{(1)}+c^{(2)}+\cdots+c^{(i)}$ becomes  less than $L(c^{(i)})$, our key approach is try to construct a cube $c_1^{(i)}$, so that
 $L(c_1^{(i)})=L(c^{(i)})$ and the linear complexity of $c^{(0)}+c^{(1)}+c^{(2)}+\cdots+c_1^{(i)}$ becomes 0 by changing $k^{(i+1)}$ elements, which implies that $c^{(0)}+c^{(1)}+c^{(2)}+\cdots+c_1^{(i)}$ has exactly $k^{(i+1)}$ nonzero elements.

Therefore the computation of  $k^{(i+1)}$ is equivalent to counting the nonzero elements of $c^{(0)}+c^{(1)}+c^{(2)}+\cdots+c_1^{(i)}$.

In the  principle of inclusion-exclusion, if $A_1\cap A_2$ is not empty, then $|A_1\cup A_2|=|A_1|+ |A_2| -|A_1\cap A_2|$.

In the computation of  $k^{(i+1)}$, if $W_H(S(c^{(0)})\cap S(c^{(1)}))\ne0$, then $c^{(0)}$ and $c^{(1)}$ can have common nonzero elements, the number of  nonzero elements of $c^{(0)}+c^{(1)}$ can become $W_H(c^{(0)})+W_H(c^{(1)})-2\times 2^{S(c^{(0)})\cap S(c^{(1)})}$.

\

From Theorem 3.1, suppose that  $s^{(n)}=c^{(0)}+c^{(1)}+c^{(2)}+t^{(n)}$, where $c^{(0)}$ is a cube with  linear complexity $L^{}(s^{(n)})$,     $c^{(1)}$ is a cube with linear complexity $L^{(1)}(s^{(n)})$,    $c^{(2)}$ is a cube with linear complexity $L^{(2)}(s^{(n)})$, and
$L(t^{(n)})<L^{(2)}(s^{(n)})$.

By Theorem 3.2,  $k^{(2)}
=2^{W_H(c^{(0)})}+2^{W_H(c^{(1)})}-2\times2^{W_H(S(c^{(0)})\cap S(c^{(1)}) )}$. After changing  $k^{(2)}$ nonzero elements, the sequence  becomes $c^{(2)}+t^{(n)}$, where $L(t^{(n)})<L(c^{(2)})$. Thus $W_H(c^{(0)}+c^{(1)})=2^{W_H(c^{(0)})}+2^{W_H(c^{(1)})}-2\times2^{W_H(S(c^{(0)})\cap S(c^{(1)}) )}$.

Without loss of generality, we consider the  superposition  of $c^{(0)}$ and $c^{(1)}$ with the alignment of first nonzero elements of two cubes. Then  $c^{(0)}+c^{(1)}$ has exactly $k^{(2)}
=2^{W_H(c^{(0)})}+2^{W_H(c^{(1)})}-2\times2^{W_H(S(c^{(0)})\cap S(c^{(1)}) )}$
  nonzero elements.

We construct a cube $c_1^{(2)}$   with linear complexity $L^{(2)}(s^{(n)})$, and furthermore, we consider the  superposition  of $c^{(0)}$, $c^{(1)}$ and $c_1^{(2)}$ with the alignment of first nonzero elements of three cubes. Then
with an analysis similar to the  principle of inclusion-exclusion, we have that
 $c^{(0)}+c^{(1)}+c_1^{(2)}$ has exactly $2^{W_H(S(c^{(0)}))}+2^{W_H(S(c^{(1)}))}+2^{W_H(S(c^{(2)}))}-2\times2^{W_H(S(c^{(0)})\cap S(c^{(1)}))}-2\times2^{W_H(S(c^{(0)})\cap S(c^{(2)}) )}-2\times2^{W_H(S(c^{(1)})\cap S(c^{(2)}))}+4\times2^{W_H(S(c^{(0)})\cap S(c^{(1)})\cap S(c^{(2)}))}$
  nonzero elements.

By adding $c^{(0)}+c^{(1)}+c_1^{(2)}$ to  $s^{(n)}=c^{(0)}+c^{(1)}+c^{(2)}+t^{(n)}$, we have $c_1^{(2)}+ c^{(2)}+t^{(n)}$.  From Lemma 2.2, $L(c_1^{(2)}+ c^{(2)})<L^{(2)}(s^{(n)})$.
Thus $k^{(3)}\le W_H(c^{(0)}+c^{(1)}+c_1^{(2)})=2^{W_H(S(c^{(0)}))}+2^{W_H(S(c^{(1)}))}+2^{W_H(S(c^{(2)}))}-2\times2^{W_H(S(c^{(0)})\cap S(c^{(1)}))}-2\times2^{W_H(S(c^{(0)})\cap S(c^{(2)}) )}-2\times2^{W_H(S(c^{(1)})\cap S(c^{(2)}))}+4\times2^{W_H(S(c^{(0)})\cap S(c^{(1)})\cap S(c^{(2)}))}$

(For example, let \\
$c^{(0)}=\{11000000\ 11000000\ 00000000\ 00000000\ 00000000\ 00000000\ 00000000\ 00000000\}$,
\\
$c^{(1)}=\{10101010\  00000000\ 10101010\  00000000\ 00000000\ 00000000\ 00000000\ 00000000 \}$,
\\
$c^{(2)}=\{11110000\ 00000000\ 00000000\ 00000000\ 11110000\
00000000\ 00000000\ 00000000\}$.

Then $c^{(0)}+c^{(1)}+c^{(2)}$ has exactly $2^{W_H(S(c^{(0)}))}+2^{W_H(S(c^{(1)}))}+2^{W_H(S(c^{(2)}))}-2\times2^{W_H(S(c^{(0)})\cap S(c^{(1)}))}-2\times2^{W_H(S(c^{(0)})\cap S(c^{(2)}) )}-2\times2^{W_H(S(c^{(1)})\cap S(c^{(2)}))}+4\times2^{W_H(S(c^{(0)})\cap S(c^{(1)})\cap S(c^{(2)}))}=4+8+8-2-4-4+4=14$
  nonzero elements.)

In the case that   $W_H[S^{(1)}(s^{(n)})\cap (S^{(0)}(s^{(n)})\cup S^{(2)}(s^{(n)}))]<W_H( S^{(1)}(s^{(n)}))$ and
$\{S^{(0)}(s^{(n)})\cap S^{(2)}(s^{(n)})=S^{(0)}(s^{(n)})\cap S^{(1)}(s^{(n)})\cap S^{(2)}(s^{(n)})$ or $[i_{m(S1\setminus S0S2)} >\min\{S^{-1}(S^{(1)}(s^{(n)})\cap S^{(2)}(s^{(n)})),S^{-1}(S^{(0)}(s^{(n)})\cap S^{(2)}(s^{(n)}))\}$ and $(S_{>i_{m(S1\setminus S0S2)}}^{(0)}(s^{(n)})\cap S_{>i_{m(S1\setminus S0S2)}}^{(2)}(s^{(n)}))\subset S^{(1)}(s^{(n)})]\}$,  we try to construct a cube $c_{-1}^{(2)}$ with  linear complexity $L(c_{}^{(2)})$, so that $c^{(0)}+c^{(1)}+c_{-1}^{(2)}$ has
less nonzero elements than $c^{(0)}+c^{(1)}+c^{(2)}$.

As   $W_H[S^{(1)}(s^{(n)})\cap (S^{(0)}(s^{(n)})\cup
S^{(2)}(s^{(n)}))]<W_H( S^{(1)}(s^{(n)}))$, there exist
$2^{W_H(S(c^{(0)})\cap S(c^{(1)})\cap S(c^{(2)}))}$   nonzero
elements in $c^{(1)}$, so that such nonzero elements  will not be
canceled by addition operation with $c^{(0)}$ or $c^{(2)}$.

In the case that $\{S^{(0)}(s^{(n)})\cap
S^{(2)}(s^{(n)})=S^{(0)}(s^{(n)})\cap S^{(1)}(s^{(n)})\cap
S^{(2)}(s^{(n)})$ or\\ $[i_{m(S1\setminus S0S2)}
>\min\{S^{-1}(S^{(1)}(s^{(n)})\cap
S^{(2)}(s^{(n)})),S^{-1}(S^{(0)}(s^{(n)})\cap S^{(2)}(s^{(n)}))\}$
and\\ $(S_{>i_{m(S1\setminus S0S2)}}^{(0)}(s^{(n)})\cap
S_{>i_{m(S1\setminus S0S2)}}^{(2)}(s^{(n)}))\subset
S^{(1)}(s^{(n)})]\}$, one can move the first $2^{W_H(S(c^{(0)})\cap
S(c^{(1)})\cap S(c^{(2)}))}$   nonzero elements in $c^{(2)}$ to the
corresponding locations in which the nonzero elements only appear in
$c^{(1)}$. In this case, $2\times2^{W_H(S(c^{(0)})\cap
S(c^{(1)})\cap S(c^{(2)}))}$ additional  nonzero elements will be
cancelled in $c^{(0)}+c^{(1)}+c_{-1}^{(2)}$, where  $c_{-1}^{(2)}$
is the new cube with  linear complexity $L(c_{}^{(2)})$.

(We follow the above example, let,\\
 $c_{-1}^{(2)}=\{01111000\ 00000000\ 00000000\ 00000000\ 01111000\ 00000000\ 00000000\ 00000000\}$.

 Then $c^{(0)}+c^{(1)}+c_{-1}^{(2)}$ has $4+8+8-2-2-4=12$ nonzero elements.)

\

In other cases, if we move the first $2^{W_H(S(c^{(0)})\cap
S(c^{(1)})\cap S(c^{(2)}))}$   nonzero elements in $c^{(2)}$
similarly as above, one can find that nonzero elements will not be
reduced after adding operation of these three sequences.

(For example, let \\
$c^{(0)}=\{10100000\ 10100000\ 00000000\ 00000000\ 00000000\ 00000000\ 00000000\ 00000000\}$,
\\
$c^{(1)}=\{11001100\  00000000\ 11001100\  00000000\ 00000000\ 00000000\ 00000000\ 00000000 \}$,
\\
$c^{(2)}=\{10101010\ 00000000\ 10101010\ 00000000\ 10101010\ 00000000\ 10101010\ 00000000\}$.

Then $S(c^{(0)})=\{ 001010\}$, $S(c^{(1)})=\{ 010101\}$,$S(c^{(2)})=\{ 110110\}$. \\
 $W_H[(S(c^{(0)})\cap S(c^{(1)}))\cup(S(c^{(1)})\cap S(c^{(2)}))]=2<W_H( S(c^{(1)}))=3$  but\\
$S(c^{(0)})\cap S(c^{(2)})=\{ 000010\}\supset S(c^{(0)})\cap S(c^{(1)})\cap S(c^{(2)})=\{ 000000\}$.

As  $S(c^{(0)})\cap S(c^{(2)})=\{ 000010\}$, $S(c^{(1)})\cap S(c^{(2)})=\{ 010100\}$, $S(c^{(1)})\setminus [S(c^{(1)})\cap (S(c^{(0)})\cup S(c^{(2)}))]=\{ 000001\}$,
so $S^{-m}(\{ 000001\})=1<S^{-1}(\{ 000010\})=2<S^{-1}(\{ 010100\})=4$.

Assume that $c_{-1}^{(2)}=\{01010101\ 00000000\ 01010101\ 00000000\ 01010101\ 00000000\ 01010101\ 00000000\}$.
 Then $c^{(0)}+c^{(1)}+c_{-1}^{(2)}$ still has $4+8+16-2-8=18$ nonzero elements.)

This completes the proof.

\end{proof}\

Next we give some examples in different situations to illustrate the effectiveness of Theorem 3.3.

\noindent {\bf Example 3.1} Let \\
$c^{(0)}=\{10001000\ 00000000\ 00000000\ 00000000\ 00000000\ 00000000\ 00000000\ 00000000\}$,
\\
$c^{(1)}=\{11000000\  11000000\ 11000000\  11000000\ 00000000\ 00000000\ 00000000\ 00000000 \}$,
\\
$c^{(2)}=\{11111111\ 00000000\ 11111111\ 00000000\ 11111111\ 00000000\ 11111111\ 00000000\}$.

Then $S(c^{(0)})=\{ 000100\}$, $S(c^{(1)})=\{ 011001\}$,$S(c^{(2)})=\{ 110111\}$. \\
 $W_H[(S(c^{(0)})\cap S(c^{(1)}))\cup(S(c^{(1)})\cap S(c^{(2)}))]=2<W_H( S(c^{(1)}))=3$  and\\
$S(c^{(0)})\cap S(c^{(2)})=\{ 000100\}\supset S(c^{(0)})\cap S(c^{(1)})\cap S(c^{(2)})=\{ 000000\}$.

As $S(c^{(1)})\cap S(c^{(2)})=\{ 010001\}$, $S(c^{(1)})\setminus [S(c^{(1)})\cap (S(c^{(0)})\cup S(c^{(2)}))]=\{ 001000\}$,
so $i_{m(S1\setminus S0S2)}=S^{-m}(\{ 001000\})=8>S^{-1}(\{ 010001\})=1$.
As $S_{>i_{m(S1\setminus S0S2)}}(c^{(0)})\cap S_{>i_{m(S1\setminus S0S2)}}(c^{(2)})=\{ 000000\}$,
thus this is the case that $[i_{m(S1\setminus S0S2)} >\min\{S^{-1}(S^{(1)}(s^{(n)})\cap S^{(2)}(s^{(n)})),S^{-1}(S^{(0)}(s^{(n)})\cap S^{(2)}(s^{(n)}))\}$ and $(S_{>i_{m(S1\setminus S0S2)}}^{(0)}(s^{(n)})\cap S_{>i_{m(S1\setminus S0S2)}}^{(2)}(s^{(n)}))\subset S^{(1)}(s^{(n)})]\}$.

We move the first $2^{W_H(S(c^{(0)})\cap S(c^{(1)})\cap S(c^{(2)}))}$   nonzero elements in $c^{(2)}$ to the location  below the not cancelled  nonzero elements in $c^{(1)}$.
Let,\\
$c_{-1}^{(2)}=\{01111111\ 10000000\ 01111111\ 10000000\ 01111111\ 10000000\ 01111111\ 10000000\}$.

 It is obvious that $c^{(0)}+c^{(1)}+c_{-1}^{(2)}$ contains exactly $2+2^3+2^5-2\times2^0-2\times2^1-2\times2^2+2\times2^0=30$  nonzero elements.
So
$k^{(3)}=30$.

\

\noindent {\bf Example 3.2}
Let \\
$c^{(0)}=\{11001100\ 00000000\ 00000000\ 00000000\}$,
\\
$c^{(1)}=\{10101010\  10101010\  00000000\ 00000000 \}$,
\\
$c^{(2)}=\{11001100\ 11001100\ 11001100\ 11001100\}$.

Then $S(c^{(0)})=\{ 00101\}$, $S(c^{(1)})=\{ 01110\}$,$S(c^{(2)})=\{ 11101\}$. \\
 $W_H[(S(c^{(0)})\cap S(c^{(1)}))\cup(S(c^{(1)})\cap S(c^{(2)}))]=2<W_H( S(c^{(1)}))=3$  and\\
$S(c^{(0)})\cap S(c^{(2)})=\{ 00101\}\supset S(c^{(0)})\cap S(c^{(1)})\cap S(c^{(2)})=\{ 00100\}$.

As $S(c^{(1)})\cap S(c^{(2)})=\{ 01100\}$, $S(c^{(1)})\setminus [S(c^{(1)})\cap (S(c^{(0)})\cup S(c^{(2)}))]=\{ 00010\}$,
so $i_{m(S1\setminus S0S2)}=S^{-m}(\{ 00010\})=2>S^{-1}(\{ 00101\})=1$.
As $S_{>i_{m(S1\setminus S0S2)}}(c^{(0)})\cap S_{>i_{m(S1\setminus S0S2)}}(c^{(2)})=\{ 00100\}\subset S(c^{(1)})$,
thus this is the case that $[i_{m(S1\setminus S0S2)} >\min\{S^{-1}(S^{(1)}(s^{(n)})\cap S^{(2)}(s^{(n)})),S^{-1}(S^{(0)}(s^{(n)})\cap S^{(2)}(s^{(n)}))\}$ and $(S_{>i_{m(S1\setminus S0S2)}}^{(0)}(s^{(n)})\cap S_{>i_{m(S1\setminus S0S2)}}^{(2)}(s^{(n)}))\subset S^{(1)}(s^{(n)})]\}$.

We move the first $2^{W_H(S(c^{(0)})\cap S(c^{(1)})\cap S(c^{(2)}))}$   nonzero elements in $c^{(2)}$ to the location  below the not cancelled  nonzero elements in $c^{(1)}$.
Let,\\
$c_{-1}^{(2)}=\{01100110\ 01100110\ 01100110\ 01100110\}$.

 It is obvious that $c^{(0)}+c^{(1)}+c_{-1}^{(2)}$ contains exactly $2^2+2^3+2^4-2\times2^1-2\times2^2-2\times2^2+2\times2^1=12$  nonzero elements.
So
$k^{(3)}=12$.

\

\noindent {\bf Example 3.3}
 Let \\
$c^{(0)}=\{11110000\ 00000000\ 00000000\ 00000000\ 00000000\ 00000000\ 00000000\ 00000000\}$,
\\
$c^{(1)}=\{11111111\ 11111111\ 00000000\ 00000000\ 00000000\ 00000000\ 00000000\ 00000000 \}$,
\\
$c^{(2)}=\{11111111\ 00000000\ 11111111\ 00000000\ 11111111\ 00000000\ 11111111\ 00000000\}$.

Then $S(c^{(0)})=\{ 000011\}$, $S(c^{(1)})=\{ 001111\}$,$S(c^{(2)})=\{ 110111\}$. \\
 $W_H[(S(c^{(0)})\cap S(c^{(1)}))\cup(S(c^{(1)})\cap S(c^{(2)}))]=3<W_H( S(c^{(1)}))=4$  and\\
$S(c^{(0)})\cap S(c^{(2)})=S(c^{(0)})\cap S(c^{(1)})\cap S(c^{(2)})=\{ 000011\}$.

So this is the case that $S^{(0)}(s^{(n)})\cap S^{(2)}(s^{(n)})=S^{(0)}(s^{(n)})\cap S^{(1)}(s^{(n)})\cap S^{(2)}(s^{(n)})$.

We move the first $2^{W_H(S(c^{(0)})\cap S(c^{(1)})\cap S(c^{(2)}))}$   nonzero elements in $c^{(2)}$ to the location  below the not cancelled  nonzero elements in $c^{(1)}$.
Let,\\
 $c_{-1}^{(2)}=\{00001111\ 11110000\ 00001111\ 11110000\ 00001111\ 11110000\ 00001111\ 11110000 \}$,   \\
It is obvious that $c^{(0)}+c^{(1)}+c_{-1}^{(2)}$ contains exactly $2^2+2^4+2^5-2\times2^2-2\times2^2-2\times2^3+2\times2^2=28$  nonzero elements.
So
$k^{(3)}=28$.

\

For $k^{(3)}$, it is easy to verify that Proposition 3.1 is the special case of Theorem 3.3.

We have tested all $2^n$-periodic binary sequences ($n=4,5$) by a computer program to verify  Theorem 3.3.

\

Investigating the linear complexity and $k$-error linear complexity of sequences  is a popular research topic.
Based on the proposed $k$-error cube decomposition and the  inclusion-exclusion  principle, one can further study
the $i$th descent point of the $k$-error linear complexity for $i>3$.

Furthermore, the proposed constructive approach can be used to
construct $2^n$-periodic binary sequences with the given linear
complexity and $k$-error linear complexity  (or CELCS). In detail,
with a given series of linear complexity values
$L(c^{(0)}),L(c^{(1)}),L(c^{(2)}), \cdots, L(c^{(m)})$, our focus is
how to construct a sequence $s^{(n)}$ with the $k$-error cube
decomposition $s^{(n)}=c^{(0)}+c^{(1)}+c^{(2)}+\cdots+c^{(m)}$,
satisfying $L(c^{(i)})=L^{(i)}(s^{(n)})$. Now we answer the
following question, if only $L(c^{(0)}),L(c^{(1)}),L(c^{(2)}) $ are
given, how many possible sequences in such decomposition?

\section{Counting functions  for $2^n$-periodic binary
sequences with given $3$-error linear complexity as the second descent point}

Next, we will study the $k$-error linear complexity of
$2^n$-periodic binary sequences by further using the sieve approach and
Games-Chan algorithm. The adopted approach is similar to
\cite{Zhou_Liu} but different from those in
\cite{Fu,Meidl2005,Zhu}.
 The proposed constructive approach in this paper is  based on the
following framework. Let $S=\{s | L(s)=c\}, E=\{e | W_H(e)=k\},
S+E=\{s+e | s\in S, e\in E\}$, where $s$ is a sequence with linear
complexity $c$ and $e$ is  sequence  with
$W_H(e)=k$. With the following sieve method, we aim to sieve sequences $s+e$ with
$L_{k}(s+e)=c$ from $S+E$.

For given linear complexity $c$, it remains to investigate two
cases. One is that $s+u\in S+E$, but $L_{k}(s+u) <c$. This is
equivalent to checking if there exists a sequence $v$ such that
$L(u+v)=c$. The other is the case that
 $s+u, t+v\in S+E$ and $L_{k}(s+u) =L_{k}(t+v) =c$ with $s\ne t$, $u\ne v$, but $s+u= t+v$.
 It is equivalent  to checking if there exists a sequence $v$ such that $L(u+v)=L(s+t)<c$ and if so, check the number of such sequence $v$, where $W_H(u)= W_H(v)=k$.

Suppose that    $s^{(n)}$ is a $2^n$-periodic binary sequence.  We first investigate the relationship between the
 first descent point   of the $k$-error linear complexity and  the
 second descent point   of the $k$-error linear complexity.
Second, based on the first descent point  and  the
 second descent point, we obtain the complete counting
functions of $2^n$-periodic binary sequences with given first descent point $k_1$-error linear complexity and second descent point $k_2$-error linear complexity.

\noindent {\bf Theorem  4.1}  Let  $s^{(n)}$ be a $2^n$-periodic binary sequence with  linear
complexity $2^n $.  Then $L_{3}(s^{(n)})<L_{1}(s^{(n)})$ if and only if $L_{1}(s^{(n)})=2^n -(2^i+2^j), 0\le i<j<n$.

  \begin{proof} $\Rightarrow$

   By result from Kurosawa et al.  \cite{Kurosawa} we know that the minimum number $k$ for which the $k$-error linear
complexity of  $2^n$-periodic binary sequence   with  linear
complexity $2^n-(2^i+2^j) $ is strictly less
than $2^n-(2^i+2^j) $ is $2^2=4$. Note that from the sequence with linear
complexity $L_{1}(s^{(n)})$ to the sequence with linear
complexity $L_{3}(s^{(n)})$, at most 4 elements have been changed. Thus,
  if $L_{3}(s^{(n)})<L_{1}(s^{(n)})$, then $s^{(n)}$ is obtained by changing one element of  a $2^n$-periodic binary sequence with  linear
complexity $2^n-(2^i+2^j) $. So $L_{1}(s^{(n)})=2^n -(2^i+2^j)$.

  $\Leftarrow$

  Suppose that $L_{1}(s^{(n)})=2^n -(2^i+2^j)$.
  Similarly by result from Kurosawa et al.  \cite{Kurosawa} we know that it is possible to change 3 elements of $s^{(n)}$, so that the new sequence with  linear
complexity less then $2^n-(2^i+2^j) $. That is $L_{3}(s^{(n)})<L_{1}(s^{(n)})$.

\end{proof}\

Next we investigate the distribution of $L_{3}(s^{(n)})$.

\noindent {\bf Theorem  4.2}  Let  $s^{(n)}$ be a $2^n$-periodic binary sequence with  linear
complexity $2^n $.  If $L_{1}(s^{(n)})=2^n -(2^i+2^j), 0\le i<j<n$, then $L_{3}(s^{(n)})=  2^n-(2^{i_1}+2^{i_2}+\cdots+2^{i_m})<2^n -(2^i+2^j)$,
where  $0\le i_1< i_2<\cdots<i_m<n, m>2$, or  $L_{3}(s^{(n)})=  2^n-(2^{i_1}+2^{i_2})<2^n -(2^i+2^j)$, where $i_1\ne i,j$ and $  i_2\ne j$.

  \begin{proof} The following proof is based on the framework: $S+E=\{t+e | t\in S, e\in E\}$.

We only give the following example  to illustrate the proof.

Let  $s^{(4)}$ be a $2^4$-periodic binary sequence with  linear
complexity $2^4 $.  If $L_{1}(s^{(4)})=2^4 -(2^0+2)$, then $L_{3}(s^{(4)})\ne  2^4-(2+2^3)$.

We will prove it by a contradiction.
Suppose that $L_{3}(s^{(4)})=  2^4-(2+2^3)$.
Let $S=\{t | L(t)=2^4-(2+2^3)\}, E=\{e | W_H(e)=3\},
S+E=\{t+e | t\in S, e\in E\}$, where $t$ is a sequence with linear
complexity $2^4-(2+2^3)$ and $e$ is  sequence  with
$W_H(e)=3$. With the sieve method, we aim to sieve sequences $t+e$ with
$L_{3}(t+e)=2^4-(2+2^3)$ from $S+E$.

We now  investigate the case that $t+u\in S+E$, but $L_{3}(t+u) <2^4-(2+2^3)$. This is
equivalent to checking if there exists a sequence $v\in E$ such that
$L(u+v)=2^4-(2+2^3)$.

For any $u\in E$ such that  $L_{1}(t+u) =2^4-(1+2)$. Such as
$u=\{1110\ 0000\ 0000\ 0000 \}$.  There exists a sequence $v\in E$ such that
$L(u+v)=2^4-(2+2^3)$. So $L_{3}(t+u) <2^4-(2+2^3)$. Here $v=\{0100\ 0000\ 1010\ 0000 \}$.

This completes the proof.

\end{proof}\

 We next derive the counting formula of
binary sequences with both the given 1-error linear complexity and the given 3-error linear complexity.

\noindent {\bf Theorem  4.3}  Let  $s^{(n)}$ be a $2^n$-periodic binary sequence with  linear
complexity $2^n $.

1) If $L_{1}(s^{(n)})=2^n -(2^i+2^j), 0\le i<j<n$, and $L_{3}(s^{(n)})=  2^n-(2^{i_1}+2^{i_2}+\cdots+2^{i_m})<2^n -(2^i+2^j)$,
where  $0\le i_1< i_2<\cdots<i_m<n, m>2$ or  $L_{3}(s^{(n)})=  2^n-(2^{i_1}+2^{i_2})<2^n -(2^i+2^j)$, where $i_1\ne i,j$ and $  i_2\ne j$.
Then the
number of $2^n$-periodic binary sequences  $s^{(n)}$ can be given by
$$2^{3n-j-i-3}\times 2^{L-1}/(2^{\epsilon+j-i_0}\times8^{n-i_m-1}) $$
where $i_0\le j$ is the minimum number  for which $2^n-(2^{i_0}+2^{j})< 2^n-(2^{i_1}+2^{i_2}+\cdots+2^{i_m})$ with a default choice  $i_0=j$.
Further, if  $j=i_m$ or $2^n-(2^{j}+2^{i_m})>L_{3}(s^{(n)})$ then $\epsilon=0$, if  $j<i_m$ and only $2^n-(2^{j}+2^{i_m})<L_{3}(s^{(n)})$ then $\epsilon=1$, if  $2^n-(2^{i}+2^{i_m})<L_{3}(s^{(n)})$ then $\epsilon=2$, where
$i_m=i_2$ for $L=2^n-(2^{i_1}+2^{i_2})$.

2) If $L_{3}(s^{(n)})=0$, then the
number of $2^n$-periodic binary sequences  $s^{(n)}$ can be given by
$2^{3n-j-i-3}$.

\begin{proof}\ 1)
Let $S=\{t | L(t)=L\}, E=\{e | W_H(e)=3\},
S+E=\{t+e | t\in S, e\in E\}$, where $t$ is a sequence with linear
complexity $L=2^n-(2^{i_1}+2^{i_2}+\cdots+2^{i_m})$ and $e$ is  sequence  with
$W_H(e)=3$ and $L_{1}(e) =2^n-(2^i+2^j)$. With the sieve method, we aim to sieve sequences $t+e$ with
$L_{3}(t+e)=L$ from $S+E$.

By Lemma 2.4, we know that
 the number of $2^n$-periodic binary sequences  $t$ with  $L(t)=L$ is $2^{L-1}$. Now we will obtain  the number of sequences $e$ with
$W_H(e)=3$ and $L_{1}(e) =2^n-(2^i+2^j)$.

Suppose that $s^{(i)}$ is a $2^{i}$-periodic binary sequence with linear complexity  $2^{i}$ and $W_H(s^{(i)})=1$,
then  the number of these $s^{(i)}$ is $2^{i}$.

So the number of $2^{i+1}$-periodic binary sequences $s^{(i+1)}$ with linear complexity $2^{i+1}-2^{i}=2^{i}$ and $W_H(s^{(i+1)})=2$ is also $2^{i}$.

For $j>i$,
if $2^{j}$-periodic binary sequences $s^{(j)}$ with linear complexity $2^{j}-2^{i}$ and $W_H(s^{(j)})=2$,
then $2^{j}-2^{i}-(2^{i+1}-2^{i})=2^{j-1}+2^{j-2}+\cdots+2^{i+1}$.

 Based on Algorithm 2.1,
the number of these $s^{(j)}$ can be given by
$(2^2)^{j-i-1}\times2^{i}=2^{2j-i-2}$.

(For example, suppose that $i=1, j=3$, then there are
$(2^2)^{j-i-1}=4$ sequences of $s^{(j)}$ correspond to a sequence
\{1010\} of $s^{(i+1)}$, given by

\{1010\ 0000\}, \{1000\ 0010\}, \{0010\ 1000\}, \{0000\ 1010\})

So the number of $2^{j+1}$-periodic binary sequences $s^{(j+1)}$ with linear complexity $2^{j+1}-(2^{j}+2^{i})$ and $W_H(s^{(j+1)})=4$ is also $2^{2j-i-2}$.

As $u\in E$ such that  $L_{1}(u) =2^n-(2^i+2^j)$. So the number of these $u$ can be given by
$$2^2\times(2^3)^{n-j-1}\times2^{2j-i-2}=2^{3n-j-i-3}$$

 We now investigate the case that
 $s+u, t+v\in S+E$ and $L_{3}(s+u) =L_{3}(t+v) =L$ with $s\ne t$, $u\ne v$, but $s+u= t+v$.
 It is equivalent  to checking if there exists a sequence $v$ such that $L(u+v)=L(s+t)<L$ and if so,
 check the number of such sequence $v$, where $W_H(u)= W_H(v)=3$.
 We need to consider the following two cases.

The first case is related to the minimum $i_0\le j$ such that $2^n-(2^{i_0}+2^j)<L= 2^n-(2^{i_1}+2^{i_2}+\cdots+2^{i_m})$.
 For any $u\in E$,  it is easy to show that there exist $2^{j-i_0}-1$ sequences $v$, such that $L(u+v)<L$.

(The following example is given to illustrate the above case.

 Suppose that $n=5, i=0,j=4, i_0=2$, $i_1=0, i_2=1, i_3=4$. So $L= 2^n-(2^{i_1}+2^{i_2}+2^{i_3})=13$.

  If
$u^{(5)}=\{1100\ 0000\ 0000\ 0000\ 1000\ 0000\ 0000\ 0000 \}$.
Then

 $v_1^{(5)}=\{0100\ 0000\ 1000\ 0000\ 0000\ 0000\ 1000 \  0000\}$,

$v_2^{(5)}=\{0100\ 1000\ 0000\ 0000\ 0000\ 1000\ 0000\ 0000 \}$,

$v_3^{(5)}=\{0100\ 0000\ 0000\ 1000\  0000\ 0000\ 0000\ 1000 \}$.

Thus $L(u^{(5)}+v_1^{(5)})= 2^5-(2^3+2^4)$, $L(u^{(5)}+v_2^{(5)})= L(u^{(5)}+v_3^{(5)})=2^5-(2^2+2^4)$.
)

\

The second case is related to $i_m<w<n$.
For $i_m<w<n$, there exist $7\times 8^{w-i_m-1}$ sequences $v$, such that $L(u+v)=2^n-(2^{i}+2^{w})<L$ or $L(u+v)=2^n-(2^{j}+2^{w})<L$ or $L(u+v)=2^n-2^{w}<L$.

Note that for any sequence $v$ with 3 nonzero elements, if we double the period of sequence $v$, then $2^3$ new sequences will be generated.
Therefore
there exist $$ 7+7\times8+\cdots+7\times8^{n-i_m-2}=8^{n-i_m-1}-1$$ sequences $v$, such that $L(u+v)<L$.

(The following example is given to illustrate the above case.

 Suppose that $n=5, i=0,j=1, i_1=1, i_2=2, i_3=3$,

$u^{(5)}=\{1110\ 0000\ 0000\ 0000\ 0000\ 0000\ 0000\ 0000 \}$.
Then

 $v_1^{(5)}=\{0100\ 0000\ 0000\ 0000\ 1010\ 0000\ 0000\ 0000 \}$,

$v_2^{(5)}=\{1000\ 0000\ 0000\ 0000\ 0110\ 0000\ 0000\ 0000 \}$,

 $v_3^{(5)}=\{0010\ 0000\ 0000\ 0000\ 1100\ 0000\ 0000\ 0000 \}$,

 $v_4^{(5)}=\{0110\ 0000\ 0000\ 0000\ 1000\ 0000\ 0000\ 0000 \}$,

 $v_5^{(5)}=\{1010\ 0000\ 0000\ 0000\ 0100\ 0000\ 0000\ 0000 \}$,

 $v_6^{(5)}=\{1100\ 0000\ 0000\ 0000\ 0010\ 0000\ 0000\ 0000 \}$,

 $v_7^{(5)}=\{0000\ 0000\ 0000\ 0000\ 1110\ 0000\ 0000\ 0000 \}$.

Thus $L(u^{(5)}+v_1^{(5)})= 2^5-(2+2^4)$, $L(u^{(5)}+v_2^{(5)})= L(u^{(5)}+v_3^{(5)})=2^5-(1+2^4)$,
$L(u^{(5)}+v_4^{(5)})= L(u^{(5)}+v_5^{(5)})=L(u^{(5)}+v_6^{(5)})= L(u^{(5)}+v_7^{(5)})=2^5-2^4$.
)

If  $j<i_m$ and only $2^n-(2^{j}+2^{i_m})<L$ then
the number of $v$ will be increased by $8^{n-i_m-1}$.

If  $2^n-(2^{i}+2^{i_m})<L$ then the number of $v$ will be increased by $3\times8^{n-i_m-1}$.

It follows that
 the
number of $2^n$-periodic binary sequences  $s^{(n)}$ with  $L_{1}(s^{(n)})=2^n -(2^i+2^j)$ and  $L_{3}(s^{(n)})=  L$
can be given by
$$2^{3n-j-i-3}\times 2^{L-1}/(2^{\epsilon+j-i_0}\times8^{n-i_m-1}) $$
where if $j=i_m$ or $2^n-(2^{j}+2^{i_m})>L$ then $\epsilon=0$, if only $2^n-(2^{j}+2^{i_m})<L$ then $\epsilon=1$, if  $2^n-(2^{i}+2^{i_m})<L$ then $\epsilon=2$.

If $j>i_0$, then $2^n-(2^{i_0}+2^j)<2^n-(2^{i_1}+2^{i_2}+\cdots+2^{i_m})<2^n-(2^{i}+2^{j})$, so $j=i_m$. If $\epsilon>0$, then $j<i_m$. Therefore, $j-i_0$ and $\epsilon$
can not be positive at the same time.

\

 We can use almost the same method to deal with the case of $L_{3}(s^{(n)})=  2^n-(2^{i_1}+2^{i_2})$ but without the situation of $j=i_2$.

 2) This is an obvious case.

\end{proof}\

To further illustrate Theorem  4.3,   we give the following two examples, which are  verified by computer program as well.

\noindent {\bf Example 4.1} Suppose that $n=4, i=1,j=3, i_0=2, i_1=0, i_2=1, i_3=3$. Note that  $L=2^n-(2^{i_1}+2^{i_2}+2^{i_3})=2^4-(1+2+8)=5$, so
$2^n-(2^{i_0}+2^j)=2^4-(4+8) <L$. As $j=i_3$, so $\epsilon=0$. The
number of $2^4$-periodic binary sequences  $s^{(4)}$ with  $L_{1}(s^{(4)})=6$ and  $L_{3}(s^{(4)})=  5$
can be given by $$ 2^{3\times n -3-1-3}\times2^{5-1}/(2^1\times 8^{4-3-1})=2^8$$

\noindent {\bf Example 4.2} Suppose that $n=4, i=1,j=2,  i_1=0, i_2=3$. Note that  $L=2^n-(2^{i_1}+2^{i_2})=2^4-(1+8)=7$. As
 $j<i_2$, $2^n-(2^{j}+2^{i_2})=4<L$ and $2^n-(2^{i}+2^{i_2})=6<L$, so $\epsilon=2$. The
number of $2^4$-periodic binary sequences  $s^{(4)}$ with $L_{1}(s^{(4)})=10$ and  $L_{3}(s^{(4)})= 7$
can be given by $$ 2^{3\times n -2-1-3}\times2^{7-1}/(2^2\times 8^{4-3-1})=2^{10}$$

\section{Counting functions  for $2^n$-periodic binary
sequences with given $4$-error linear complexity as the second descent point}

Next  based on the sieve approach and cube theory,  we derive the counting formula of  $2^n$-periodic
binary sequences with both the given 2-error linear complexity as the first descent point and 4-error linear complexity as the second descent point.

In the case of given $3$-error linear complexity, as  $L_{}(s^{(n)})$ is a constant, the counting formula of  $2^n$-periodic
binary sequences basically has two parameters: $L_{1}(s^{(n)})$ and $L_{3}(s^{(n)})$.
In the case of given $4$-error linear complexity, the counting formula of  $2^n$-periodic
binary sequences basically has three parameters: $L_{}(s^{(n)})$, $L_{2}(s^{(n)})$ and $L_{4}(s^{(n)})$. So this is a much complicated case.

\

It is known by result from Kurosawa et al.  \cite{Kurosawa} that for a $2^n$-periodic binary sequence with  linear
complexity $2^n -(2^i+2^j), 0\le i<j<n$, $4$-error linear complexity is the first descent point.  However, with cube theory we will characterize
 $2^n$-periodic binary sequences with  $4$-error linear complexity as the second descent point.

\noindent {\bf Theorem  5.1}  Let  $s^{(n)}$ be a $2^n$-periodic binary sequence with  linear
complexity less than $2^n $.  Then

i). Suppose that $c_1$ and $c_2$ are in the standard cube decomposition of sequence $s^{(n)}$ and $L(s^{(n)})=L(c_1)$. If $L_{4}(s^{(n)})<L_{2}(s^{(n)})<L(s^{(n)})$,
then $c_1$ and $c_2$ are two 1-cubes or $c_1$ is a  1-cube and $c_2$ is a 2-cube;

ii).  $L_{4}(s^{(n)})<L_{2}(s^{(n)})<L(s^{(n)})$ if and only if  $L_{2}(s^{(n)})=2^n -(2^i+2^j), 0\le i<j<n$, but $L_{2}(s^{(n)})\ne2^n -(1+2)$;

iii).  If  $L_{}(s^{(n)})=2^n -2^{i_0}$, then $i_0<i$ or $i<i_0<j$.
  \begin{proof}\  i).
 Suppose that $s^{(n)}$ is a $2^n$-periodic binary sequence with  linear
complexity $2^n-(2^{i_1}+2^{i_2}+\cdots+2^{i_m})$. By Kurosawa et al.  \cite{Kurosawa} we know that the minimum number $k$ for which the $k$-error linear
complexity of  $2^n$-periodic binary sequence $s^{(n)}$ is strictly less
than the linear complexity  of $s^{(n)}$ is $2^m$. So the proof  is obvious.

ii).  Based on i), here we only need to prove that $L_{2}(s^{(n)})\ne2^n -(1+2)$.

In the case that $c_1$ and $c_2$ are two 1-cubes.
As $L(s^{(n)})\ne2^n -(1+2)$, there exist two nonzero elements with distance $d>2$ in $c_1$ and $c_2$. Suppose that $L_{2}(s^{(n)})=2^n -(2^i+2^j)$. Then $2^j\ge d>2$. It follows that $L_{2}(s^{(n)})\ne2^n -(1+2)$.

In the case that $c_1$ is a  1-cube and $c_2$ is a 2-cube.
If $L(c_2)=2^n -(1+2)$, then $L(c_1)=2^n -1$ or $2^n -2$. There exist two nonzero elements with distance $d>2$ in $c_1$ and $c_2$. Suppose that $L_{2}(s^{(n)})=2^n -(2^i+2^j)$. Then $2^j\ge d>2$. It follows that $L_{2}(s^{(n)})\ne2^n -(1+2)$.

iii). Based on i) and ii), it is easy to prove iii).
\end{proof}\

Next we investigate the distribution of $L_{4}(s^{(n)})$.

\noindent {\bf Theorem  5.2}  Let  $s^{(n)}$ be a $2^n$-periodic binary sequence with  linear
complexity  $L(s^{(n)})=2^n-2^{i_0} $. If  $L_{4}(s^{(n)})<L_{2}(s^{(n)})<L(s^{(n)})$
and $L_{2}(s^{(n)})=2^n -(2^i+2^j), 0\le i<j<n$, then $L_{4}(s^{(n)})=  2^n-(2^{i_1}+2^{i_2}+\cdots+2^{i_m})<2^n -(2^i+2^j)$,
where  $0\le i_1< i_2<\cdots<i_m<n, m>3$, or $L_{4}(s^{(n)})=  2^n-(2^{i_1}+2^{i_2}+2^{i_3})$, where  $\{i_1,i_2,i_3\}\ne \{i,j,i_0\}$, $\{i_1,i_2,i_3\}\ne \{0,1,2\}$, or
  $L_{4}(s^{(n)})=  2^n-(2^{i_1}+2^{i_2})<2^n -(2^i+2^j)$, where $i_2\ne j$, $i_1\ne i,j,i_0$.

  \begin{proof} The following proof is based on the framework: $S+E=\{t+e | t\in S, e\in E\}$.

In the case that  $L_{4}(s^{(n)})=  2^n-(2^{i_1}+2^{i_2}+\cdots+2^{i_m})<2^n -(2^i+2^j)$, the proof is obvious.

In the case that  $L_{4}(s^{(n)})=  2^n-(2^{i_1}+2^{i_2})<2^n -(2^i+2^j)$.
We only give the following example  to illustrate the proof.

Let  $s^{(4)}$ be a $2^4$-periodic binary sequence with  linear
complexity less than $2^4 $.  If $L_{}(s^{(4)})=2^4 -2$ and $L_{2}(s^{(4)})=2^4 -(1+2^2)$, then $L_{4}(s^{(4)})\ne  2^4-(2+2^2)$.

Suppose that $L_{4}(s^{(4)})=  2^4-(2+2^2)$.
Let $S=\{t | L(t)=2^4-(2+2^2)\}, E=\{e | W_H(e)=4\},
S+E=\{t+e | t\in S, e\in E\}$, where $t$ is a sequence with linear
complexity $2^4-(2+2^2)$ and $e$ is  sequence  with
$W_H(e)=4$. With the sieve method, we aim to sieve sequences $t+e$ with
$L_{4}(t+e)=2^4-(2+2^2)$ from $S+E$.

We now  investigate the case that $s+u\in S+E$, but $L_{4}(t+u) <2^4-(2+2^2)$. This is
equivalent to checking if there exists a sequence $v\in E$ such that
$L(u+v)=2^4-(2+2^2)$.

For any $u\in E$ such that  $L_{2}(t+u) =2^4-(1+4)$. Such as
$u=\{1100\ 0110\ 0000\ 0000 \}$.  There exists a sequence $v\in E$ such that
$L(u+v)=2^4-(2+2^2)$. So $L_{4}(t+u) <2^4-(2+2^2)$. Here
$v=\{1001\ 0011\ 0000\ 0000 \}$ such that  $L_{2}(t+v) =2^4-(1+4)$. Therefore $i_2\ne j$.

Let $L(t)=2^4-(1+2^3)$. There exists a sequence $v\in E$ such that
$L(u+v)=2^4-(1+2^3)$. So $L_{4}(t+u) <2^4-(1+2^3)$. Here
$v=\{0000\ 0110\ 1100\ 0000 \}$ such that  $L_{2}(t+v) =2^4-(1+2^2)$. Therefore $i_1\ne i$.

Let $L(t)=2^4-(2^2+2^3)$. There exists a sequence $v\in E$ such that
$L(u+v)=2^4-(2^2+2^3)$. So $L_{4}(t+u) <2^4-(2^2+2^3)$. Here
$v=\{1000\ 0010\ 0100\ 0100 \}$ such that  $L_{2}(t+v) =2^4-(1+2^2)$. Therefore $i_1\ne j$.

Let $L(t)=2^4-(2+2^3)$. There exists a sequence $v\in E$ such that
$L(u+v)=2^4-(2+2^3)$. So $L_{4}(t+u) <2^4-(2+2^3)$. Here
$v=\{0100\ 0100\ 1000\ 0010 \}$ such that  $L_{2}(t+v) =2^4-(1+2^2)$. Therefore $i_1\ne i_0$.

\

In the case that  $L_{4}(s^{(n)})=  2^n-(2^{i_1}+2^{i_2}+2^{i_3})$. Note that $c_1$ is a  1-cube and $c_2$ is a 2-cube,  $L(c_2)=L_{2}(s^{(n)})=2^n -(2^i+2^j)$.
We also use the following example  to illustrate the  proof.

 Suppose that $n=4, i=0,j=3, i_0=2$,
$u^{(4)}=\{0100\ 1000\ 1100\ 0000 \}$.

Then there exists
$v^{(4)}=\{1000\ 0100\ 0000\ 1100  \}$, such that $L(u^{(4)}+v^{(4)})=2^4-(1+2^2+2^3)$. Therefore $L_{4}(s^{(n)})\ne  2^n-(2^{i_1}+2^{i_2}+2^{i_3})$, where  $\{i_1,i_2,i_3\}= \{i,i_0,j\}$.

Now we consider the case that $L_{4}(s^{(n)})=  2^n-(2^{0}+2^{1}+2^{2})$. As $L_{4}(s^{(n)})<2^n -(2^i+2^j)<2^n -2^{i_0}$, so $2^{i_0}<2^i+2^j<2^{0}+2^{1}+2^{2}$.
Suppose that $L(t)=  2^n-(2^{0}+2^{1}+2^{2})$. For any $u\in E$ such that  $L_{2}(t+u) =2^n-(2^i+2^j)$. It is easy to prove that $L_{4}(t+u)< 2^n-(2^{0}+2^{1}+2^{2})$.
We just use the following example  to illustrate the  proof.

 Suppose that $n=4, i=0,j=2, i_0=1$,
$u^{(4)}=\{0110\ 1100\ 0000\ 0000 \}$ and $t^{(4)}=\{1111\ 1111\ 0000\ 0000 \}$, then $t^{(4)}+u^{(4)}=\{1001\ 0011\ 0000\ 0000 \}$.
So, $L_4(t^{(4)}+u^{(4)})=0$.

If $t^{(4)}=\{1111\ 0011\ 0000\ 1100 \}$, then $L_4(t^{(4)}+u^{(4)})=2^4-(2^{0}+2^{3})$.

This completes the proof.
\end{proof}

\

 We next derive the counting formula of
binary sequences with both the given 2-error linear complexity and the given 4-error linear complexity.

\noindent {\bf Theorem  5.3}  Let  $s^{(n)}$ be a $2^n$-periodic binary sequence with  linear
complexity  $L(s^{(n)})=2^n-2^{i_0} $.

1) If  $L_{4}(s^{(n)})<L_{2}(s^{(n)})<L(s^{(n)})$ and $L_{2}(s^{(n)})=2^n -(2^i+2^j), 0\le i<j<n$, and $L_{4}(s^{(n)})=  2^n-(2^{i_1}+2^{i_2}+\cdots+2^{i_m})<2^n -(2^i+2^j)$,
where  $0\le i_1< i_2<\cdots<i_m<n, m>3$ or  $L_{4}(s^{(n)})=  2^n-(2^{i_1}+2^{i_2}+2^{i_3})$, where $\{i_1,i_2,i_3\}\ne \{i,j,i_0\}$, $\{i_1,i_2,i_3\}\ne \{0,1,2\}$, or
  $L_{4}(s^{(n)})=  2^n-(2^{i_1}+2^{i_2})<2^n -(2^i+2^j)$, where $i_2\ne j$, $i_1\ne i,j,i_0$.
Then the
number of $2^n$-periodic binary sequences  $s^{(n)}$ can be given by
$$(2^{4n-j-i-4-i_0}/\gamma)\times 2^{L-1}/(2^{\delta+\epsilon}\times16^{n-i_m-1}) $$
where if $i_0>i $ then $\gamma=2$ else $\gamma=1$; if  $2^n-(2^{i}+2^{i_0}+2^j)>L$ then $\delta=0$, if only $2^n-(2^{i}+2^{i_0}+2^j)<L$ then $\delta=1$,
 if  $2^n-(2^{i_0}+2^j)<L$ then $\delta=2$. Further, if $j=i_m$ or $2^n-(2^{j}+2^{i_m})>L$ then $\epsilon=0$,
 if  $j<i_m$ and only $2^n-(2^{j}+2^{i_m})<L$ then $\epsilon=1$, if  $2^n-(2^{i}+2^{i_m})<L$ and  $2^n-(2^{i_0}+2^{i_m})>L$ then $\epsilon=2$, if  $2^n-(2^{i}+2^{i_m})>L$ and  $2^n-(2^{i_0}+2^{i_m})<L$ then $\epsilon=2$, if  $2^n-(2^{i}+2^{i_m})<L$ and  $2^n-(2^{i_0}+2^{i_m})<L$ then $\epsilon=3$, where
$i_m=i_3$ for $L=2^n-(2^{i_1}+2^{i_2}+2^{i_3})$ and $i_m=i_2$ for $ L=2^n-(2^{i_1}+2^{i_2})$.

2) If $L_{4}(s^{(n)})=0$, then the
number of $2^n$-periodic binary sequences  $s^{(n)}$ can be given by
$2^{4n-j-i-4-i_0}/\gamma$.

\

\begin{proof}\ 1)
Let $S=\{t | L(t)=L\}, E=\{e | W_H(e)=4\},
S+E=\{t+e | t\in S, e\in E\}$, where $t$ is a sequence with linear
complexity $L=2^n-(2^{i_1}+2^{i_2}+\cdots+2^{i_m}), m>2$ and $e$ is  sequence  with
$W_H(e)=4$ and $L_{2}(e) =2^n-(2^i+2^j)$. With the sieve method, we aim to sieve sequences $t+e$ with
$L_{4}(t+e)=L$ from $S+E$.

By Lemma 2.4, we know that
 the number of $2^n$-periodic binary sequences  $t$ with  $L(t)=L$ is $2^{L-1}$.
 Now we will compute  the number of sequences $e$ with
$W_H(e)=4$ and $L_{2}(e) =2^n-(2^i+2^j)$.

Suppose that $s^{(i)}$ is a $2^{i}$-periodic binary sequence with linear complexity  $2^{i}$ and $W_H(s^{(i)})=1$,
then  the number of these $s^{(i)}$ is $2^{i}$

So the number of $2^{i+1}$-periodic binary sequences $s^{(i+1)}$ with linear complexity $2^{i+1}-2^{i}=2^{i}$ and $W_H(s^{(i+1)})=2$ is also $2^{i}$.

For $j>i$,
if $2^{j}$-periodic binary sequences $s^{(j)}$ with linear complexity $2^{j}-2^{i}$ and $W_H(s^{(j)})=2$,
then $2^{j}-2^{i}-(2^{i+1}-2^{i})=2^{j-1}+2^{j-2}+\cdots+2^{i+1}$.

 Based on Algorithm 2.1,
the number of these $s^{(j)}$ can be given by
$(2^2)^{j-i-1}\times2^{i}=2^{2j-i-2}$.

So the number of $2^{j+1}$-periodic binary sequences $s^{(j+1)}$ with linear complexity $2^{j+1}-(2^{j}+2^{i})$ and $W_H(s^{(j+1)})=4$ is also $2^{2j-i-2}$.

As $u\in E$ such that  $L_{2}(u) =2^n-(2^i+2^j)$. So the number of these $u$ can be given by
$$2^2\times\frac{2^{j+1}}{2^{i_0+1}\times \gamma} \times(2^4)^{n-j-1}\times2^{2j-i-2}=2^{4n-j-i-4-i_0}/\gamma$$
where if $i_0>i $ then $\gamma=2$ else $\gamma=1$.

( The following example is given to illustrate the  case of $i_0>i $.

 Suppose that $n=4, i=0,j=3, i_0=2$,
$u^{(4)}=\{0100\ 0000\ 1100\ 1000 \}$.

Then it may comes from
 $v_1^{(4)}=\{1100\ 0000\ 1100\ 0000 \}$,
or $v_2^{(4)}=\{0100\ 1000\ 0100\ 1000 \}$,
where $L_2(v_1^{(4)})=L_2(v_2^{(4)})= 2^4-(1+2^3)$.
)

\

 We now investigate the case that
 $s+u, t+v\in S+E$ and $L_{4}(s+u) =L_{4}(t+v) =L$ with $s\ne t$, $u\ne v$, but $s+u= t+v$.
 It is equivalent  to checking if there exists a sequence $v$ such that $L(u+v)=L(s+t)<L$ and if so,
 check the number of such sequence $v$, where $W_H(u)= W_H(v)=4$.
 We need to consider the following two cases.

The first case is related to $i_0$. For any $u\in E$, there exists one sequence $v$, such that $L(u+v)=2^n-(2^{i}+2^{i_0}+2^j)<L$, and there exist two sequences $v$, such that $L(u+v)=2^n-(2^{i_0}+2^j)<L$.

( The following example is given to illustrate the above case.

 Suppose that $n=4, i=1,j=3, i_0=2$,
$u^{(4)}=\{1000\ 0010\  1010\ 0000 \}$.
Then

 $v_1^{(4)}=\{0010\ 1000\ 0000\ 1010 \}$,

$v_2^{(4)}=\{0000\ 1010\ 0010\ 1000 \}$,

 $v_3^{(4)}=\{1010\ 0000\ 1000\ 0010 \}$.

Thus $L(u^{(4)}+v_1^{(4)})= 2^4-(2+2^2+2^3)$, $L(u^{(4)}+v_2^{(4)})= L(u^{(4)}+v_3^{(4)})=2^4-(2^2+2^3)$.
)

\

The second case is related to $i_m<w<n$. For $i_m<w<n$, there exist $15\times 16^{w-i_m-1}$ sequences $v$,
such that $L(u+v)=2^n-(2^{i}+2^{w})<L$ or $L(u+v)=2^n-(2^{j}+2^{w})<L$
or $L(u+v)=2^n-(2^{i_0}+2^{w})<L$ or $L(u+v)=2^n-2^{w}<L$.

Note that for any sequence $v$ with 4 nonzero elements, if we double the period of sequence $v$, then $2^4$ new sequences will be generated.
Therefore
there exist $$ 15+15\times16+\cdots+15\times16^{n-i_m-2}=16^{n-i_m-1}-1$$ sequences $v$, such that $L(u+v)<L$.

( The following example is given to illustrate the above case.

 Suppose that $n=5, i=0,j=2, i_0=1, i_1=1, i_2=2, i_3=3, w=4$,

$u^{(5)}=\{1001\ 1100\ 0000\ 0000\ 0000\ 0000\ 0000\ 0000 \}$.
Then

 $v_1^{(5)}=\{0001\ 0100\ 0000\ 0000\ 1000\ 1000\ 0000\ 0000 \}$.

  Thus $L(u^{(5)}+v_1^{(5)})= 2^5-(2^2+2^4)$.

 $v_2^{(5)}=\{1000\ 1000\ 0000\ 0000\ 0001\ 0100\ 0000\ 0000 \}$,

$v_{3}^{(5)}=\{0000\ 0000\ 0000\ 0000\ 1001\ 1100\ 0000\ 0000  \}$.

Thus $L(u^{(5)}+v_2^{(5)})= L(u^{(5)}+v_3^{(5)})=2^5-(2+2^4)$.

$v_4^{(5)}=\{1001\ 0000\ 0000\ 0000\ 0000\ 1100\ 0000\ 0000 \}$,

$v_5^{(5)}=\{0000\ 1100\ 0000\ 0000\ 1001\ 0000\ 0000\ 0000  \}$,

 $v_6^{(5)}=\{0001\ 1000\ 0000\ 0000\ 1000\ 0100\ 0000\ 0000 \}$,

 $v_7^{(5)}=\{ 1000\ 0100\ 0000\ 0000\ 0001\ 1000\ 0000\ 0000 \}$.

Thus $L(u^{(5)}+v_4^{(5)})=L(u^{(5)}+v_5^{(5)})=L(u^{(5)}+v_6^{(5)})= L(u^{(5)}+v_7^{(5)})=2^5-(1+2^4)$.

$\cdots\cdots$

$v_{15}^{(5)}=\{1000\ 0000\ 0000\ 0000\ 0001\ 1100\ 0000\ 0000  \}$.

Thus $L(u^{(5)}+v_{15}^{(5)})= 2^5-2^4$.
)

\

On the other hand, if $j<i_m$ and only $2^n-(2^{j}+2^{i_m})<L$ then
the number of $v$ will be increased by $16^{n-i_m-1}$.

If  $2^n-(2^{i}+2^{i_m})<L$ and   $2^n-(2^{i_0}+2^{i_m})>L$ then the number of $v$ will be increased by $3\times16^{n-i_m-1}$.

If  $2^n-(2^{i}+2^{i_m})>L$ and   $2^n-(2^{i_0}+2^{i_m})<L$ then the number of $v$ will be increased by $3\times16^{n-i_m-1}$.

If  $2^n-(2^{i}+2^{i_m})<L$ and   $2^n-(2^{i_0}+2^{i_m})<L$ then the number of $v$ will be increased by $7\times16^{n-i_m-1}$.

\

It follows that
 the
number of $2^n$-periodic binary sequences  $s^{(n)}$ with $L(s^{(n)})=2^n-2^{i_0} $,  $L_{2}(s^{(n)})=2^n -(2^i+2^j)$ and  $L_{4}(s^{(n)})=  L$
can be given by
$$(2^{4n-j-i-4-i_0}/\gamma)\times 2^{L-1}/(2^\delta\times 2^\epsilon\times16^{n-i_m-1}) $$
where if  $2^n-(2^{i}+2^{i_0}+2^j)>L$ then $\delta=0$, if only $2^n-(2^{i}+2^{i_0}+2^j)<L$ then $\delta=1$, if  $2^n-(2^{i_0}+2^j)<L$ then $\delta=2$;
if $j=i_m$ or $2^n-(2^{j}+2^{i_m})>L$ then $\epsilon=0$, if $j<i_m$ and only $2^n-(2^{j}+2^{i_m})<L$ then $\epsilon=1$, if  $2^n-(2^{i}+2^{i_m})<L$
and $2^n-(2^{i_0}+2^{i_m})>L$ then $\epsilon=2$, if  $2^n-(2^{i}+2^{i_m})>L$ and $2^n-(2^{i_0}+2^{i_m})<L$ then $\epsilon=2$, if  $2^n-(2^{i}+2^{i_m})<L$ and $2^n-(2^{i_0}+2^{i_m})<L$ then $\epsilon=3$.

If $\delta>0$, then $2^n-(2^{i}+2^{i_0}+2^j)< 2^n-(2^{i_1}+2^{i_2}+\cdots+2^{i_m})<2^n-(2^{i}+2^{j})$, so $j=i_m$. If $\epsilon>0$, then $j<i_m$. Therefore, $\delta$ and $\epsilon$
can not be positive at the same time.

\

 We can use almost the same method to deal with the case of $L_{4}(s^{(n)})=  2^n-(2^{i_1}+2^{i_2})$ but without the situation of $j=i_2$.

  2) This is an obvious case.
\end{proof}

\

To further illustrate Theorem  5.3,   we give the following two examples, which are  verified by computer program as well.

\noindent {\bf Example 5.1} Suppose that $n=4, i=1,j=3, i_0=2, i_1=0, i_2=1, i_3=3$. Note that $i_0>i$, so $\gamma=2$. As $L=2^n-(2^{i_1}+2^{i_2}+2^{i_3})=2^4-(1+2+8)=5$, so
$2^n-(2^{i}+2^{i_0}+2^j)=2^4-(2+4+8) <L$ and $2^n-(2^{i_0}+2^j)=2^4-(4+8) <L$. Thus $\delta=2$. As $j=i_3$, so $\epsilon=0$. The
number of $2^4$-periodic binary sequences  $s^{(4)}$ with $L(s^{(4)})=12 $,  $L_{2}(s^{(4)})=6$ and  $L_{4}(s^{(4)})=  5$
can be given by $$ (2^{4\times n -3-1-4-2}/2)\times2^{5-1}/(2^2\times 16^{4-3-1})=2^7$$

\noindent {\bf Example 5.2} Suppose that $n=5, i=2,j=3, i_0=1, i_1=0, i_2=4$. Note that $i_0<i$, so $\gamma=1$. As $L=2^n-(2^{i_1}+2^{i_2})=2^5-(1+16)=15$, so
$2^n-(2^{i}+2^{i_0}+2^j)=2^5-(4+2+8) >L$. Thus $\delta=0$. As $j<i_2$  and  $2^n-(2^{i}+2^{i_2})=12 <L$ and $2^n-(2^{i_0}+2^{i_2})=14<L$ so $\epsilon=3$. The
number of $2^5$-periodic binary sequences  $s^{(5)}$ with $L(s^{(5)})=30 $,  $L_{2}(s^{(5)})=20$ and  $L_{4}(s^{(5)})=  15$
can be given by $$ 2^{4\times n -3-2-4-1}\times2^{15-1}/(2^3\times 16^{5-4-1})=2^{21}$$

\

In previous research, investigators mainly focus on the linear complexity
and $k$-error complexity for a given sequence.
As demonstrated here,   based on {\it Cube Theory}, the sieve approach
 can be used to construct
$2^n$-periodic binary sequences with the given linear complexity and
$k$-error linear complexity (or CELCS), and this is a more
challenging problem with broad applications.

\section{Conclusions}


In this paper, we first propose the  $k$-error cube decomposition
for  $2^n$-periodic binary sequences.  By applying the famous
inclusion-exclusion principle, we obtain the complete
characterization of $i$th descent point (critical point) of the
k-error linear complexity for $i=2,3$.
Second,
 via the sieve method and Games-Chan algorithm,  the
 second descent point (critical point) distribution of the $k$-error linear complexity for $2^n$-periodic binary
sequences is characterized. As a consequence, we obtain the complete counting
functions on the $k$-error linear complexity of
$2^n$-periodic binary sequences as the  second descent point  for $k=3,4$.

Furthermore, the proposed constructive approach can be used
to construct $2^n$-periodic binary sequences with the given linear complexity and $k$-error linear complexity  (or CELCS). This is a
challenging problem with broad applications.
 We will continue this work in future
due to its importance.

 \section*{ Acknowledgment}
 The research was partially supported by
Anhui Natural Science Foundation(No.1208085MF106).

\end{document}